\documentclass[final,5p,times,twocolumn]{elsarticle}

\RequirePackage{color}\definecolor{BLUE}{rgb}{0,0.20,0.75} 

\usepackage{multirow}

\usepackage{graphicx}
\usepackage{amssymb}

\usepackage{color}

\usepackage{url}

\usepackage{breakurl}
\usepackage[breaklinks]{hyperref}
\usepackage{verbatim}
\usepackage{subfigure}

\usepackage[table]{xcolor}
\usepackage{hhline}
\definecolor{tsigc}{rgb}{0.65882353, 0.65882353, 1.}
\definecolor{tnsigc}{rgb}{0.96862745, 0.53333333, 0.54901961}
\newcommand{\tsig}{\cellcolor{tsigc}}
\newcommand{\tnsig}{\cellcolor{tnsigc}}

\journal{Computer Speech and Language}

\begin{document}

\begin{frontmatter}


\title{Investigation of learning abilities on linguistic features in sequence-to-sequence text-to-speech synthesis}



\author[nii,sokendai]{Yusuke Yasuda\corref{cor}}\ead{yasuda@nii.ac.jp}
\author[nii]{Xin Wang}\ead{wangxin@nii.ac.jp}
\author[nii,sokendai,uoe]{Junichi Yamagishi}\ead{jyamagis@nii.ac.jp}

\address[nii]{National Institute of Informatics, 2-1-2 Hitotsubashi, Chiyoda-ku, Tokyo, 101-8430, Japan} 
\address[sokendai]{SOKENDAI (The Graduate University for Advanced Studies),  Shonan Village, Hayama, Kanagawa, 240-0193, Japan}
\address[uoe]{Centre for Speech Technology Research, University of Edinburgh, 10 Crichton Street
Edinburgh, EH8 9AB, UK}

\cortext[cor]{Corresponding author}

\begin{abstract}

Neural sequence-to-sequence text-to-speech synthesis (TTS) can produce high-quality speech directly from text or simple linguistic features such as phonemes. Unlike traditional pipeline TTS, the neural sequence-to-sequence TTS does not require manually annotated and complicated linguistic features such as part-of-speech tags and syntactic structures for system training. However, it must be carefully designed and well optimized so that it can implicitly extract useful linguistic features from the input features.

In this paper we investigate under what conditions the neural sequence-to-sequence TTS can work well in Japanese and English along with comparisons with deep neural network (DNN) based pipeline TTS systems. Unlike past comparative studies, the pipeline systems also use neural autoregressive (AR) probabilistic modeling and a neural vocoder in the same way as the sequence-to-sequence systems do for a fair and deep analysis in this paper. 
We investigated systems from three aspects: a) model architecture, b) model parameter size, and c) language. For the model architecture aspect, we adopt modified Tacotron systems that we previously proposed and their variants using an encoder from Tacotron or Tacotron2. For the model parameter size aspect, we investigate two model parameter sizes. For the language aspect, we conduct listening tests in both Japanese and English to see if our findings can be generalized across languages. 

Our experiments on Japanese demonstrated that the Tacotron TTS systems with increased parameter size and input of phonemes and accentual type labels outperformed the DNN-based pipeline systems using the complicated linguistic features and that its encoder could learn to compensate for a lack of rich linguistic features. Our experiments on English demonstrated that, when using a suitable encoder, the Tacotron TTS system with characters as input can disambiguate pronunciations and produce natural speech as good as those of the systems using phonemes. However, we also found that the encoder could not learn English stressed syllables from characters perfectly and hence resulted in flatter fundamental frequency.
In summary, these experimental results suggest that a) a neural sequence-to-sequence TTS system should have a sufficient number of model parameters to produce high quality speech, b) it should also use a powerful encoder when it takes characters as inputs, and c) the encoder still has a room for improvement and needs to have an improved architecture to learn supra-segmental features more appropriately.  

\end{abstract}


\end{frontmatter}


\section{Introduction}
\label{sec:intro}
Traditional text-to-speech synthesis (TTS) such as the deep neural network (DNN)-based statistical parametric speech synthesis (SPSS) framework \cite{zen2013statistical} converts input text into output waveforms by using modules in a pipeline: a text analyzer to derive linguistic features such as syntactic and prosodic tags from text, a duration model to predict the phoneme duration, an acoustic model to predict the acoustic features such as mel-cepstral coefficients and $F_0$, and a vocoder to produce the waveform from the acoustic features. Such a TTS system can produce reasonably good waveforms, but training the modules in the pipeline can be laborious. For example, the text analyzer requires manually annotated prosodic tags; the duration and acoustic models need alignment between the linguistic and acoustic feature sequences. 

Sequence-to-sequence neural TTS is a recently developed framework that uses a single model to conduct the task of all or most modules in the SPSS-based TTS pipeline. 
For example, an ideal end-to-end sequence-to-sequence TTS uses a single neural network to directly convert the text into the waveform. Because such a TTS system is expected to implicitly learn the word pronunciation, prosodic realization, speaking styles, and the alignment between text and speech, it can be trained from many TTS databases with only the waveform and the text transcription. 
In practice, a sequence-to-sequence TTS system may still leverage a separate neural waveform generator and a grapheme-to-phoneme converter for ideographic languages such as Japanese and Chinese. Although it is not fully end-to-end, such a TTS system requires no alignment and simplifies the training process.


Several sequence-to-sequence based TTS methods have been proposed so far \cite{Wang2017, Sotelo2017Char2wavES, DBLP:conf/iclr/TaigmanWPN18, DBLP:conf/iclr/PingPGAKNRM18, Shen2017, DBLP:conf/aaai/Li0LZL19}. Some of them have been evaluated against the DNN-based SPSS pipeline and unit-selection-based TTS systems, and it is widely agreed now that sequence-to-sequence based TTS can generally generate more natural synthetic speech. Particularly, Tacotron2 \cite{Shen2017}, which is a successor to Tacotron \cite{Wang2017}, and a Transformer-based TTS system \cite{DBLP:conf/aaai/Li0LZL19} have advanced the sequence-to-sequence based TTS methods to the human-level naturalness.


We applied Tacotron \cite{Wang2017} to Japanese in our previous research \cite{DBLP:conf/icassp/YasudaWTY19}. Unlike English, Japanese is not a language to which sequence-to-sequence based TTS methods can be applied straightforwardly, mainly due to two issues: character diversity and pitch accent. Our modified Tacotron systems successfully produced synthetic speech with correct pitch accent by using accentual type labels along with phonemes as inputs. However, the synthetic speech of our proposed systems did not match the naturalness of those from comparable pipeline systems using an AR acoustic model and neural vocoder. 

One possible way to 
fill the gap between our proposed systems and the pipeline systems is to introduce richer full-context label: the super set of phonemes and accentual type labels containing more complex features (e.g. part-of-speech tag and syntactic analysis). However, this strategy is the exact opposite of  end-to-end sequence-to-sequence TTS strategies. 
Another approach is empowering sequence-to-sequence TTS models themselves to compensate for their lack of complex features by implicitly extracting more from simple inputs like texts. However, little known about what kind of linguistic features can be compensated for by a powerful model and which changes (e.g.\ parameter size, neural network structure, and input linguistic feature) contribute to the improvement to what extent. This is what we want to investigate in this paper. 


In this paper, we improve our sequence-to-sequence TTS models
by following the changes made by Tacotron2 \cite{Shen2017} and compare the effect of each group of the changes. Concretely, the configuration changes include increasing the model parameter size and simplifying the encoder's network structure. Furthermore, we apply the best configurations to English TTS systems using character input and analyze if the systems can implicitly learn linguistic features such as phone and lexical stress. 

We also compared our sequence-to-sequence TTS models with the new configuration against strong pipeline systems with autoregressive (AR) probabilistic models and WaveNet-based waveform generation. Note that, in many studies, the sequence-to-sequence based methods have been compared with pipeline systems using a non-AR acoustic model, a conventional vocoder for waveform generation, or both \cite{Wang2017, Shen2017, DBLP:conf/iclr/PingPGAKNRM18}. Our comparison may be fairer because recent studies have shown that AR modeling is crucial for not only sequence-to-sequence \cite{Watts2019} but also the DNN-based pipeline TTS systems \cite{XinWang2018}.

This paper is organized as follows. In section \ref{sec:background} we review the background of the pipeline based TTS method, sequence-to-sequence TTS methods, and our previous work about Tacotron based Japanese TTS methods. In addition, we summarize the transition from Tacotron to Tacotron2 to provide background for our new experimental conditions. In section \ref{sec:proposed-methods}, we describe our TTS systems used for this investigation. In section \ref{sec:experiment} we explain the new experimental conditions and their results. Section \ref{sec:conclusion} concludes our findings\footnote{This paper is partially based on our previous work published in \cite{DBLP:conf/icassp/YasudaWTY19}. The main focus of this journal paper is an analysis of implicit learning abilities on linguistic features in sequence-to-sequence text-to-speech synthesis in Japanese and English and differs from that of previous work, where we proposed a new neural network architecture for Japanese sequence-to-sequence text-to-speech synthesis.}. 

\section{Background}
\label{sec:background}

\subsection{TTS frameworks and Terminology}

\begin{figure}[t]
	\centering
		\includegraphics[width=1\columnwidth]{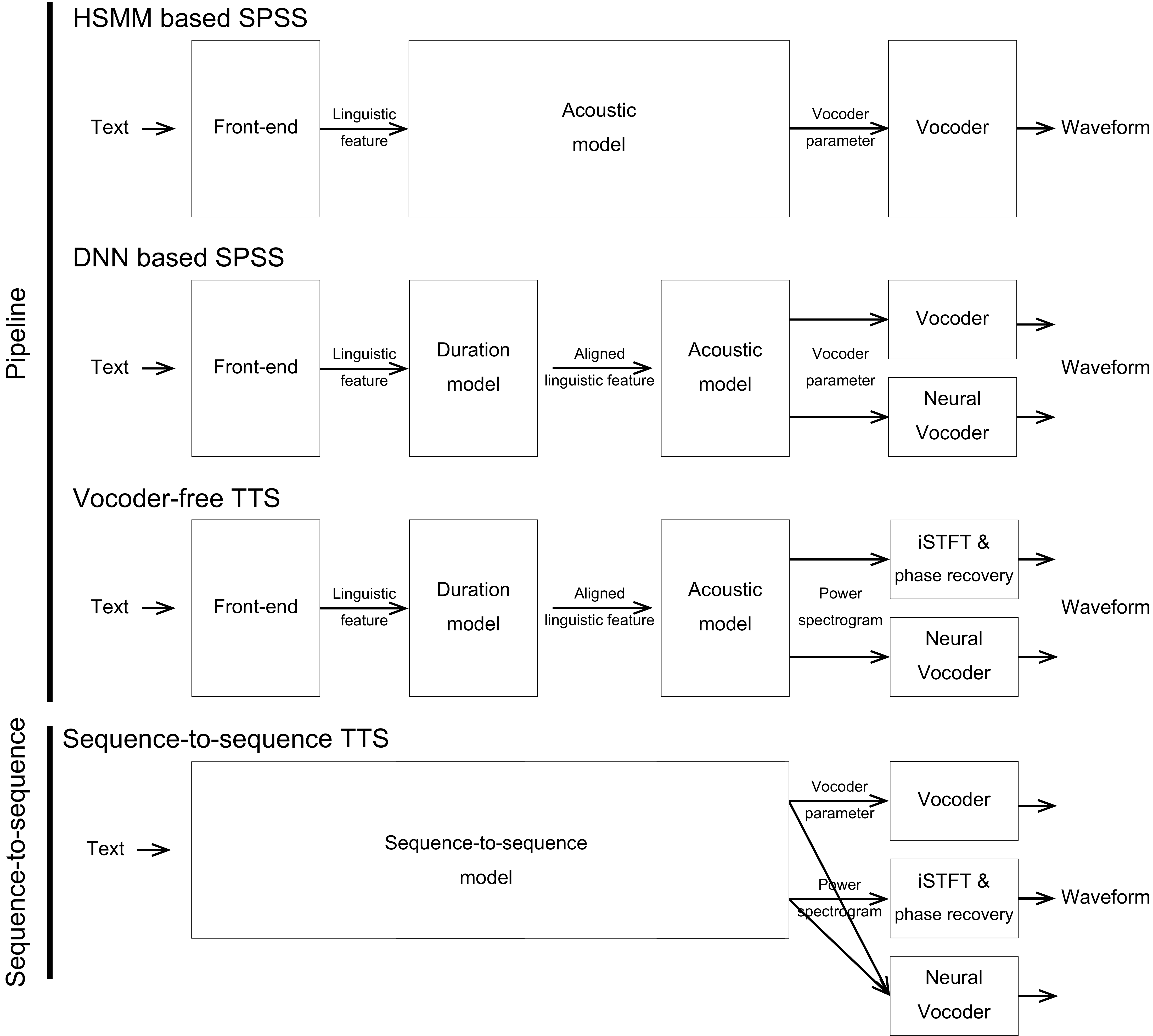}
	\caption{TTS frameworks appearing in this paper. Pipeline TTS framework consists of independent front-end and back-end models. Pipeline TTS framework includes HSMM based SPSS, DNN based SPSS, and vocoder-free TTS. HSMM based SPSS and DNN based SPSS predict vocoder parameter with HSMM or DNN, respectively, and they use conventional vocoder for waveform generation. Vocoder-free TTS predicts power spectrogram with DNN, and it uses invert STFT and phase recovery to generate waveform. Neural vocoder can also be used for waveform generation. Sequence-to-sequence TTS has a single model, and it may use either vocoder parameter or power spectrogram, and corresponding waveform generation method or neural vocoder. }
	\label{fig:tts-frameworks}
\end{figure}

This paper describes a sequence-to-sequence TTS framework and compares it with the pipeline TTS framework. Since machine learning based speech synthesis was long studied prior to the deep learning period, this section clarifies the definition of the two TTS frameworks and a taxonomy of TTS methods belonging to the frameworks. Figure \ref{fig:tts-frameworks} shows TTS frameworks appearing in this paper.

SPSS is a TTS method that involves acoustic parametric representation of speech, i.e.\ vocoder parameter \cite{ref:Zen09}. A typical SPSS-based TTS system contains an acoustic model to predict vocoder parameters and a vocoder to produce a waveform given the predicted vocoder parameters. 
For example, hidden semi-Markov model (HSMM)-based SPSS uses the HSMM as the acoustic model to convert linguistic features into vocoder parameters \cite{DBLP:conf/interspeech/ZenTMKK04}. Such an HSMM treats the phoneme duration as hidden variable and does not require an independent aligner or a phoneme duration model. HSMM-based SPSS uses a conventional vocoder to produce the waveform \cite{ref:Kawahara99,morise2016world}, which suffers from the artifact caused by minimum phase and other assumptions in the speech production process \cite{Tamamori2017}.

DNN-based SPSS \cite{zen2013statistical} uses a DNN rather than an HSMM as the acoustic model. It outperforms the HSMM-based SPSS by switching from the state-level acoustic modeling to frame-level regression \cite{watts2016hmms}.
However, typical DNN-based SPSS relies on an external duration model to align linguistic features and target vocoder parameters, even though there has been an attempt to combine HSMM with DNN \cite{Tokuda+2016}.
Traditional DNN-based SPSS has relied on the conventional vocoder used in HSMM-based SPSS. However, since the birth of neural vocoders \cite{oord2016wavenet, mehri2016samplernn}, some DNN-based SPSS systems have adopted a neural vocoder for waveform generation. These systems can improve the quality of synthetic speech by directly modeling the waveform with the neural vocoder. 

There are also TTS approaches that do not depend on vocoder parameters. In these approaches, an acoustic model predicts a power spectrogram \cite{Takaki2017}, an acoustic feature that is obtained with a simpler analysis and is closer in form to a waveform than vocoder parameters. In this method, a waveform is generated by applying inverse Short-Time Fourier Transform (STFT) with phase recovery. This approach is normally distinguished from SPSS and is referred to as vocoder-free TTS \cite{SAITO2019347}. It inherits the pipeline structure of DNN-based SPSS, but it can alleviate buzzy sounds caused by the conventional vocoder. However, it causes another type of artifact when the prediction accuracy of power spectrogram is poor or when the hop size of the windows is too large.  

DNN-based SPSS and vocoder-free TTS are major approaches in pipeline TTS frameworks in recent works. Although the distinction between the two approaches is based on the type of acoustic feature, the difference become less important once a neural vocoder is used to generate the waveform. A neural vocoder can take arbitrary acoustic features as input. Vocoder parameters \cite{Luong2018}, fundamental frequency and frame-level linguistic features \cite{oord2016wavenet}, and mel-scale amplitude spectrograms \cite{Lorenzo-Trueba2018} have been used as input of the neural vocoder in pipeline TTS. For more details on the pipeline TTS frameworks, see Appendix \ref{sec:pipeline}.

A sequence-to-sequence TTS framework is the main focus of this paper. As figure \ref{fig:tts-frameworks} shows, a sequence-to-sequence TTS framework has a different structure from the pipeline one: it unifies the front-end, duration model, and acoustic model as one model by utilizing the DNN-based sequence-to-sequence framework that can map a source-to-target sequence with a length mismatch. As well as pipeline TTS, sequence-to-sequence TTS has variant systems using different types of acoustic features and waveform generation methods: vocoder parameter and conventional vocoder \cite{Wang+2016FirstStepTowards, DBLP:conf/iclr/TaigmanWPN18}, vocoder parameter and neural vocoder \cite{Sotelo2017Char2wavES}, mel and linear spectrogram and invert STFT with phase recovery \cite{Wang2017}, and mel or linear spectrogram and neural vocoder \cite{Shen2017}. 

In this paper, 
we use “pipeline" to refer to all methods in a pipeline framework and “DNN-based pipeline TTS" to refer to both DNN-based SPSS and vocoder-free TTS. We clearly mention a specific TTS method if the distinction is important. We use ``conventional'' or ``traditional'' for the pipeline TTS framework 
even though some methods in the pipeline framework are relatively new in TTS history and the first sequence-to-sequence TTS was proposed only a few years ago \cite{Wang+2016FirstStepTowards}.

\subsection{Sequence-to-sequence based TTS}
\label{subsec:sequence-to-sequence-tts}

\subsubsection{Overview}
\label{subsubsec:sequence-to-sequence-overview}

Sequence-to-sequence based TTS is a new TTS framework using a sequence-to-sequence learning method \cite{sutskever2014sequence}. 
The sequence-to-sequence based TTS unifies the linguistic model, acoustic model, duration model, and ultimately the waveform model from the conventional pipeline framework into a single model. 
It has provoked many studies 
because of its potential to unify all models into one and its handiness as it uses text and speech only for model training.

Many architectures have been proposed for sequence-to-sequence based TTS. Notable architecture examples are Tacotron \cite{Wang2017}, Char2Wav \cite{Sotelo2017Char2wavES}, VoiceLoop \cite{DBLP:conf/iclr/TaigmanWPN18}, DeepVoice3 \cite{DBLP:conf/iclr/PingPGAKNRM18}, DCTTS \cite{Tachibana2018}, Tacotron2 \cite{Shen2017}, and Transformer \cite{DBLP:conf/aaai/Li0LZL19}. All of them have an encoder-decoder \cite{DBLP:conf/emnlp/ChoMGBBSB14} structure, which consists of an encoder to encode linguistic input and a decoder to decode acoustic features autoregressively.
From the perspective of NN types to implement the sequence-to-sequence framework, 
those TTS architectures can be roughly classified into recurrent neural network (RNN)-based \cite{Wang2017, Sotelo2017Char2wavES, Shen2017}, convolutional neural network (CNN)-based \cite{DBLP:conf/iclr/PingPGAKNRM18, Tachibana2018}, self-attention based ones \cite{DBLP:conf/aaai/Li0LZL19} or memory buffer \cite{DBLP:conf/iclr/TaigmanWPN18}.

Note that, although sequence-to-sequence based TTS is a new framework, each key element such as phone and linguistic feature embedding \cite{watts2013unsupervised, Lu2013}, autoregressive decoder \cite{Lorenzo-Trueba2018, wangxinDARf0}, and postnet \cite{Chen2015ADeepGenerative,kaneko2017generative} has already been investigated for DNN-based pipeline TTS methods separately.  A notable difference between the sequence-to-sequence and pipeline TTS frameworks is not the elements themselves but joint training of the elements.   

\subsubsection{Linguistic modeling}
The pipeline TTS frameworks typically use grapheme-to-phoneme conversion. 
The sequence-to-sequence TTS instead uses an encoder to transform input text into a hidden representation that is supposed to encode the pronunciation and possibly prosody information.

An initial proof-of-concept of sequence-to-sequence TTS was investigated in Chinese with phoneme input \cite{Wang+2016FirstStepTowards}. Then character level input was investigated in English \cite{Wang2017, Sotelo2017Char2wavES} and was confirmed to be feasible in other alphabetical languages \cite{Park2019} and Chinese \cite{Li2019KnowledgeBased, Zhang2019}. The naturalness of synthetic speech using character input can be quite high for English \cite{Shen2017}. 

Although implicit linguistic modeling is possible when the encoder has a number of sufficient non-linear transformation layers \cite{Watts2019} and when there is a sufficient amount of training data,  even large-scale speech copora such as LibriTTS \cite{Zen2019} never match a lexicon dictionary in terms of word coverage \cite{Taylor2019}. This suggests a potential limitation of end-to-end TTS for out-of-vocabulary words and out-of-domain text, especially in languages with highly nonphonemic orthography such as English and ideographic languages such as Chinese. It is reported that phoneme input gives better results than character input \cite{Fong2019AComparisonOfLetters, Zhang2019} for sequence-to-sequence based TTS and many recent sequence-to-sequence based TTS studies avoid character input and use phonemes instead. Some studies use both character and phoneme \cite{DBLP:conf/iclr/PingPGAKNRM18, DBLP:journals/corr/abs-1905-08459} or a random mix of them \cite{DBLP:conf/icassp/KastnerSBC19}. It is also reported that pre-training an encoder to learn grapheme-to-phoneme mapping is also an effective approach  \cite{Li2019KnowledgeBased}. 
Nevertheless, the sequence-to-sequence based TTS is convenient to construct a multi-lingual model by handling various inputs with a single model, and it has been investigated with character, phoneme, and even byte input \cite{Xue2019, DBLP:journals/corr/abs-1811-09364, Li2019, Zhang2019}.

There are also attempts to use richer linguistic information as additional inputs for sequence-to-sequence based TTS for tonal languages such as Chinese \cite{Tian2019}, and pitch accent languages such as Japanese \cite{Okamoto2019}, where tone or accent and prosody information are indispensable yet hard to learn from graphemes or phonemes implicitly. However, some studies report excessive information negatively affects their results \cite{Lu2019ImplementingProsodicPhrasing, Fujimoto2019}. Simpler linguistic features such as pinyin with tone \cite{DBLP:journals/corr/abs-1911-02839}, phones with tone \cite{Zhang2018, Lu2019ImplementingProsodicPhrasing}, or phonemes with accentual type \cite{DBLP:conf/icassp/YasudaWTY19} seems to be good choices for these languages.

\subsubsection{Alignment modeling}

As another difference from DNN-based pipeline TTS, the sequence-to-sequence TTS does not use the external aligner or duration model. 
Instead, many sequence-to-sequence TTS methods use an attention mechanism \cite{Bahdanau2014} to implicitly align source and target sequences. 

There are many types of attention mechanisms. 
Basic ones include Gaussian mixture model (GMM) \cite{graves2013generating}, additive \cite{Bahdanau2014}, dot-product \cite{DBLP:conf/emnlp/LuongPM15}, and location-sensitive \cite{DBLP:conf/nips/ChorowskiBSCB15} attention mechanisms, which were designed for other tasks but applied to TTS as in \cite{Sotelo2017Char2wavES, DBLP:conf/iclr/TaigmanWPN18, Wang2017, DBLP:conf/iclr/PingPGAKNRM18, DBLP:conf/aaai/Li0LZL19, Wang+2016FirstStepTowards, Shen2017}. 
Several attention mechanisms have also been proposed specifically for TTS, where the alignment between source and target is monotonic and the target sequence is much longer than the source. 
They include forward attention \cite{Zhang2018}, stepwise monotonic attention \cite{He2019}, a mixture of logistic attention \cite{DBLP:journals/corr/abs-1906-01083}, and dynamic convolution attention \cite{DBLP:journals/corr/abs-1910-10288}, all of which use a time relative approach to enforce monotonicity.

In addition, attention regularization loss has also been proposed for robust and fast alignment learning \cite{Tachibana2018, Zhu2019}. Furthermore, instead of using an advanced attention mechanism, systems based on CNN or self-attention that enable parallel training have advanced to fast non-autoregressive inference using attention distribution distillation \cite{DBLP:journals/corr/abs-1905-08459} or duration model distillation \cite{Ren2019}. Meantime, \citet{Watts2019} found out that jointly modeled alignment by attention is not the source of the high naturalness of sequence-to-sequence based TTS.


\subsubsection{Acoustic modeling}

Typically, decoders in the sequence-to-sequence TTS adopt deep autoregressive neural networks to produce the target acoustic feature sequence on the basis of the attention and encoder outputs.  

The autoregressive structure is introduced to model its statistical dependency across frames and is very important for improving the prediction accuracy \citet{Watts2019}. However, the autoregressive modeling has a common problem called \textit{exposure bias} -- The autoregressive decoder is inclined to excessively rely on feedback acoustic features while ignoring the encoder output to predict the next acoustic feature frame. Furthermore, the inaccurate predictions accumulate across the frames and lead to alignment errors and unacceptable outputs --
To alleviate this problem, most architectures apply dropout \cite{DBLP:journals/jmlr/SrivastavaHKSS14} regularization to feedback in layers called pre-net \cite{Wang2017, DBLP:conf/iclr/PingPGAKNRM18, Shen2017, DBLP:conf/aaai/Li0LZL19}. There are also methods using a Generative Adversarial Network (GAN) \cite{Guo2019}, forward-backward decoding \cite{Zheng2019}, and teacher-student training \cite{DBLP:journals/corr/abs-1911-02839} to alleviate the exposure bias.

As for acoustic feature, mel-scale spectrogram is widely used. 
The fine-grained vocoder parameter is less commonly used since it increases the length of output sequences and make decoding more difficult.


\begin{table*}[t]
\caption{Comparison of structure and configuration between Tacotron and Tacotron2 in literature and modified Tacotron models in this study. Baseline Tacotron and Self-attention Tacotron correspond to models A and B illustrated in figure~\ref{fig:japanese-tacotron}. Numbers in bracket tuple indicate size of hidden layer(s). For modified Tacotrons, number tuples before and after ``/'' denote configuration for small and large model parameter sizes in Section~\ref{sec:experiment}, respectively. Note that CBH block in encoders of modified Tacotron can be replaced with CNN (see Section~\ref{sec:experiment}). Pho., and accen., and para. denote phoneme, accentual type, and parameter, respectively.}
\label{tbl:tacotron-tacotron2}
\small{
\begin{center}
\begin{tabular}{|c|l||l|l|p{4.3cm}|p{4.3cm}|}\hline
\multicolumn{2}{|c||}{Model} & \multicolumn{2}{c|}{Tacotron in literature} 			 & \multicolumn{2}{c|}{Modified Tacotron in this study} \\ \cline{3-6}
\multicolumn{2}{|l||}{} & Tacotron \cite{Wang2017} & Tacotron2 \cite{Shen2017} & Baseline Tacotron & Self-attention Tacotron \\ \hline\hline
\multicolumn{2}{|l||}{Vocoder} & Griffin-Lim \cite{griffin1984signal} & WaveNet \cite{oord2016wavenet} & \multicolumn{2}{l|}{WaveNet, open-source version} \\ \hline\hline
\multirow{6}{*}{\rotatebox{90}{Decoder}} & Post-net & CBHG (256) & CNN (512) & {CNN (512)} & {CNN (512)} \\ \cline{2-6}
& Self-attention  & -             &  -               &  - & Self-attention (256) / (1024)  \\ \cline{2-6}
& Decoder RNN & GRU (256, 256) & -               & {LSTM (256, 256) / (1024, 1024)} & {LSTM (256, 256) / (1024, 1024)} \\ \cline{2-6}
& Attention & Additive (256) & Location-aware (128) & Forward (256) / (128) & {Additive \& Forward (256) / (128)} \\ \cline{2-6}
& Attention RNN & GRU (256) & LSTM (1024, 1024) & {LSTM (256) / (128)} & {LSTM (256) / (128)} \\ \cline{2-6}
& Pre-net & FFN (256, 128) & FFN (256, 256) & {FFN (256, 128) / (256, 256)} & {FFN (256, 128) / (256, 256)} \\ \hline\hline
\multirow{7}{*}{\rotatebox{90}{Encoder}} &   Self-attention     & -  & - & - & {Self-attention (32) / (64)}  \\  \cline{2-6}
& \multirow{2}{*}{Encoder core} & \multirow{2}{*}{CBHG (256)} &  Bi-LSTM (512)  & {Bi-LSTM (256) / (512)} & {Bi-LSTM (256) / (512)} \\ \cline{4-6}
&         &            &  CNN  (512)  & {CBH (128) / (256)} & {CBH (128) / (256)} \\ \cline{2-6}
& \multirow{2}{*}{Pre-net} & \multirow{2}{*}{FFN (256, 128)} & \multirow{2}{*}{-} & {FFN Pho. (224, 112) / (448, 224)} & {FFN Pho. (224, 112) / (448, 224)} \\ 
&  &  & & FFN Accen. (32, 16) / (64, 32) & FFN Accen. (32, 16) / (64, 32) \\ \cline{2-6}
& \multirow{2}{*}{Embedding} & \multirow{2}{*}{(256)} & \multirow{2}{*}{(512)} & Pho. (224) / (448) & Pho. (224) / (448) \\ 
&  &  &  & Accen. (32) / (64)& Accen. (32) / (64)  \\ \hline\hline
\multicolumn{2}{|l||}{\# Para. w/o vocoder} & $6.9  \times 10^6$ & $27.3 \times 10^6$ & $11.3 \times 10^6$ / $35.8 \times 10^6$ & $11.6 \times 10^6$ / $41.6 \times 10^6$\\ \hline
\end{tabular}
\end{center}
}
\end{table*}

\subsection{Transition from Tacotron to Tacotron2}
\label{subsec:tacotron-to-tacotron2}

Next, we review the differences between Tacotron and Tacotron2 since the change from Tacotron to Tacotron2 is known to be a successful example of improving the naturalness of synthetic speech in sequence-to-sequence based TTS and since we also compare these specific architectures in our comparisons. 

Details in the transition from Tacotron to Tacotron2 can be classified into three groups of changes: 1) simplified neural network modules, 2) increased model parameter size, and 3) introduction of a neural vocoder. Table \ref{tbl:tacotron-tacotron2} summarizes the network structure and parameter sizes of Tacotron and Tacotron2. Tacotron uses an encoder that consists of pre-net \cite{Wang2017} and CBHG modules, where CBH stands for Convolution Banks, Highway networks \cite{SrivastavaGS15} and G stands for bidirectional-Gated recurrent unit (GRU) \cite{Chung2014} RNN. Tacotron2 generally has simpler network structures than Tacotron: its encoder does not have pre-net \cite{Wang2017}, and the CBH module is replaced with simple CNN layers. Unlike Tacotron, Tacotron2 does not have the decoder RNN and instead has one additional layer in the attention RNN, so Tacotron2 has two layers in the attention RNN. The CBHG module in post-net \cite{Wang2017} is also replaced with CNN layers, and the role of post-net is to improve possibly oversmoothed mel-spectrograms from the autoregressive decoder, instead of converting the scale of the spectrogram from mel into linear. Meanwhile, Tacotron2 has an advanced attention mechanism. Tacotron2 has adopted location-sensitive attention \cite{DBLP:conf/nips/ChorowskiBSCB15}, an extended version of additive attention \cite{Bahdanau2014}. Location-sensitive attention can consider location in addition to content to align the source and target, so it is suitable for speech tasks and was originally tested in speech recognition. In contrast, additive attention was proposed for machine translation.

In addition to the simplified network structure, the model parameter size is increased significantly in Tacotron2. Table \ref{tbl:tacotron-tacotron2} also compares the parameter sizes of neural network layers in Tacotron and Tacotron2. Along with the expansion of parameter size, many regularization techniques have been introduced, probably because models with huge capacity are more likely to suffer from overfitting. Dropout \cite{DBLP:journals/jmlr/SrivastavaHKSS14} is applied to every CNN layer in the encoder and post-net. Zoneout \cite{Krueger2016} regularization is applied to long short-term memory (LSTM) layers in the encoder and decoder. The L2 regularization term is added in the loss term. In contrast, in Tacotron, regularization is conducted only in the encoder pre-net and decoder pre-net through dropout.

Tacotron2 uses WaveNet \cite{oord2016wavenet} for waveform synthesis instead of the Griffin-Lim algorithm \cite{griffin1984signal}. Human-level naturalness of synthetic speech was achieved by training WaveNet with a predicted spectrogram from Tacotron2. In other words, WaveNet can correct errors in a predicted spectrogram from Tacotron2 during waveform generation in this configuration.


\subsection{Japanese sequence-to-sequence TTS}
\label{subsec:japanese-tts-tacotron}

Finally, we overview sequence-to-sequence TTS systems designed for Japanese since we use Japanese as well as English for our experiments. 

In the case of Japanese, the first issue is the diversity of input text characters, i.e., thousands of ideograms called kanji (i.e.,  logographic Chinese characters) alongside two syllabic scripts called hiragana and katakana. 
The huge number of kanji characters posts a challenge for the sequence-to-sequence TTS to learn the pronunciation. 
Another challenging yet not well recognized issue is the pitch accent estimation. Unlike the pitch accent in English, Japanese pitch accent directly affects the meaning of words and the perceived naturalness of the speech. Although the pitch accent of individual words can be encoded in a lexicon, it is affected by adjacent words in an accentual phrase, i.e., pitch accent sandhi. 

In our previous research \cite{DBLP:conf/icassp/YasudaWTY19}, we investigated the Japanese TTS using Tacotron \cite{Wang2017}. We used the phoneme as the input and focused on the issue with pitch accent. Specifically, we proposed modified Tacotron architectures to model the Japanese pitch accent, which will be detailed in section \ref{sec:proposed-methods}. Through a large-scale listening test, we found that using accentual type (i.e., pitch accent in the accentual phrase) as an additional input feature significantly improved the pitch accent and the naturalness of the generated speech for native listeners. 
However, our Tacotron system using the phoneme and the accentual type as input performed worse than DNN-based pipeline systems with an autoregressive acoustic model \cite{Luong2018, Lorenzo-Trueba2018} and WaveNet \cite{oord2016wavenet} vocoder.  This could be due to the difference between Tacotron and Tacotron2 described earlier. As far as we know, human-level naturalness of synthetic speech has not been achieved for Japanese TTS using Tacotron yet.

\section{Speech synthesis systems used in this investigation}
\label{sec:proposed-methods}

\subsection{Pipeline systems}

We used two DNN-based pipeline systems: DNN-based SPSS and vocoder-free TTS. Both pipelines use an acoustic model with autoregressive density modeling and a neural vocoder. 

\begin{figure}[t]
	\centering
		\includegraphics[width=1.0\columnwidth]{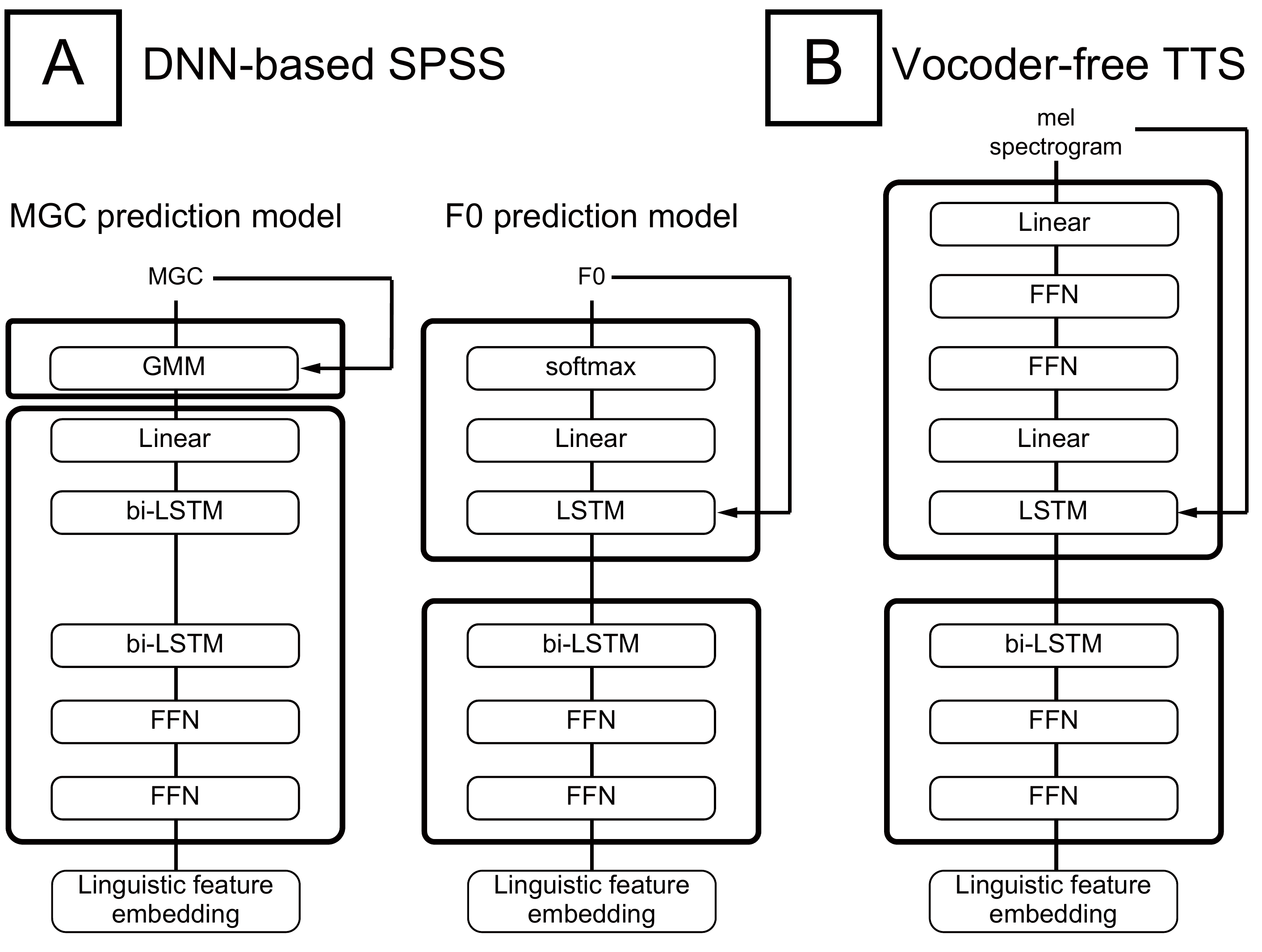}
	\caption{Acoustic models of pipeline TTS systems used in our experiments. A: DNN-based SPSS; left is MGC prediction model based on autoregressive recurrent  mixture  density  network; right is $F_0$ prediction model based on deep autoregressive recurrent neural network. B: Mel-spectrogram prediction model of Vocoder-free TTS. }
	\label{fig:pipeline-systems}
\end{figure}

\begin{figure*}[t]
	\centering
		\includegraphics[width=2\columnwidth]{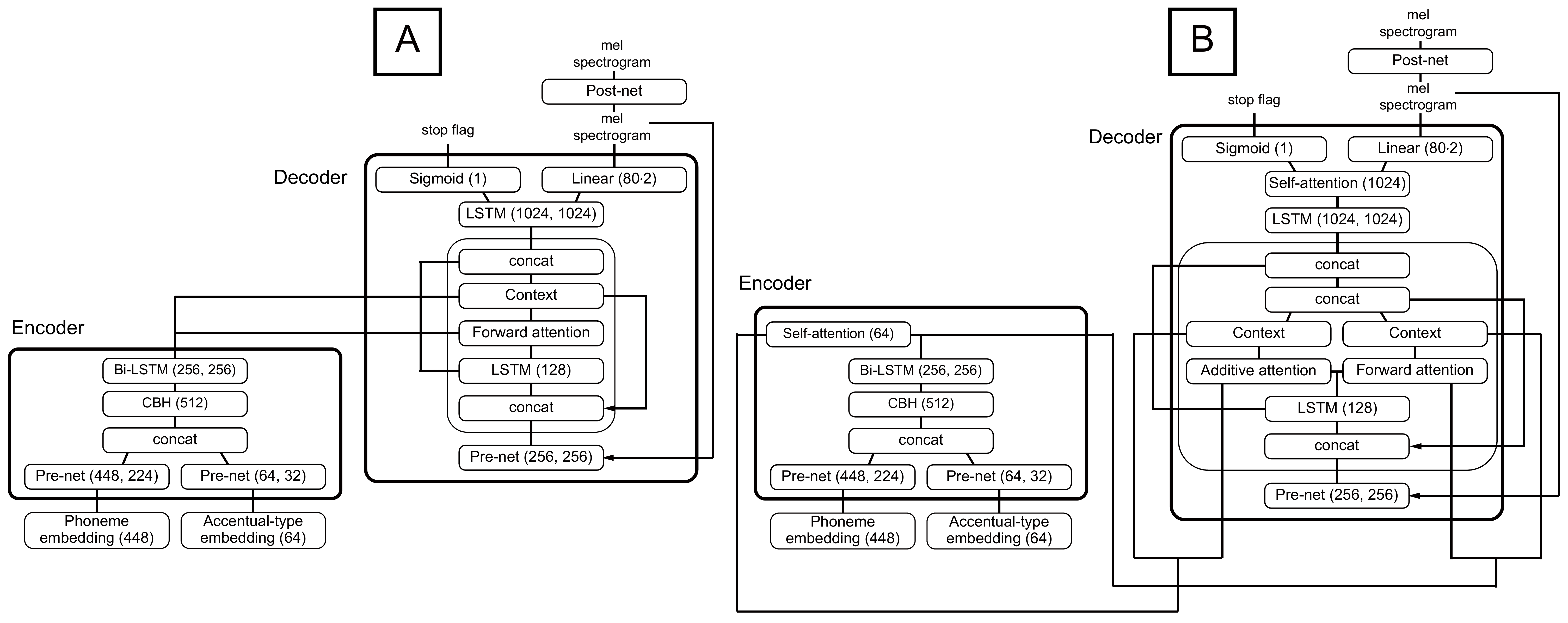}
	\caption{A: Neural network architecture of baseline Tacotron. B: Neural network architecture of self-attention Tacotron. Numbers in brackets are parameter size for layers. Arrows indicate feedback from the previous time step. Note that displayed parameter sizes are adjusted to “large" configuration for experiments conducted in this paper. Note that post-net on mel-spectrogram is added for the experiments in this study (see section \ref{subsec:condition-japanese-tts}) and was not used in the original work \cite{Wang2017, DBLP:conf/icassp/YasudaWTY19}.}
	\label{fig:japanese-tacotron}
\end{figure*}

Figure \ref{fig:pipeline-systems}-A shows the architecture of acoustic models in the DNN-based SPSS system used in this study. This system is used for Japanese TTS research \cite{Luong2018} and consists of two acoustic models to predict two vocoder parameters. The first model in the DNN-based SPSS system is responsible for modeling the mel-generalized cepstral coefficient (MGC). It is based on a bidirectional recurrent mixture density network (RMDN) \cite{DBLP:conf/nips/Schuster99}, but autoregressive feedback connection is introduced at the output layer to improve density modeling with AR \cite{wangARRMDN}. The autoregressive feedback from past adjacent frames is linearly transformed to compute output density parameters. 
The second model in the DNN-based SPSS system is a $F_0$ model. The $F_0$ output modeled in this model is discrete, based on a just-noticeable difference in pitch in speech perception \cite{moore2012introduction}. Thus, its output is a softmax distribution. This model also has also autoregressive feedback, but the feedback is nonlinearly transformed by a stack of neural network layers. 
In addition, this feedback scheme has an architecture consisting of two network modules. One module is bidirectional RNN and additional network layers that take linguistic input, and the other is a unidirectional RNN layer that runs an autoregressive loop. The former module corresponds to an encoder and the latter corresponds to a decoder in a sequence-to-sequence framework \cite{DBLP:conf/emnlp/ChoMGBBSB14}. However, unlike a sequence-to-sequence framework, this architecture's input is a frame level linguistic feature aligned by a duration model, so source and target mapping is frame-by-frame.

Figure \ref{fig:pipeline-systems}-B shows the architecture of an acoustic model in the vocoder-free TTS system used in this study. This system was used for a voice cloning study in English \cite{Lorenzo-Trueba2018}, its target acoustic feature is the mel-spectrogram. The architecture of the acoustic model consists of two network modules: a bidirectional RNN based module that processes linguistic features, and a unidirectional RNN based module with a deep autoregressive feedback scheme of nonlinear transformation, as well as the $F_0$ model in the DNN-based SPSS system.

Both the DNN-based SPSS and vocoder-free TTS systems use HSMMs to predict phoneme duration and a WaveNet neural vocoder to generate waveforms. The WaveNet predicts a quantized waveform in 10~bits with $\mu$-law coding at a 16~kHz sampling rate. 
Both systems showed high naturalness in a listening test, and their scores did not differ much \cite{DBLP:conf/icassp/YasudaWTY19}.

\subsection{Baseline Tacotron}
\label{subsec:japanese-tacotron}

We used a modified version of Tacotron \cite{DBLP:conf/icassp/YasudaWTY19} to handle pitch accent in Japanese and refer to it as baseline Tacotron. Figure \ref{fig:japanese-tacotron}-A illustrates its network structure, and Table~\ref{tbl:tacotron-tacotron2} lists its details. Our baseline Tacotron is based on Chinese Tacotron proposed by \citet{Zhang2018}. To handle pitch accent, the baseline Tacotron has separate embedding tables with different dimensions for phonemes and accentual type labels. Its encoder has two pre-nets \cite{Wang2017}: bottleneck layers for embedding vectors from phonemes and accentual type labels. The two inputs are then concatenated and encoded by CBH networks \cite{SrivastavaGS15}, and bidirectional-LSTM \cite{DBLP:journals/tsp/SchusterP97} modules. 
Note that we replace the GRU cell \cite{Chung2014} in the CBHG module with a LSTM cell to apply zoneout regularization \cite{Krueger2016}. Zoneout regularization along with the LSTM cell is introduced in the successors of the original Tacotron \cite{DBLP:conf/icml/WangSZRBSXJRS18, Shen2017}, 
and we use LSTM with the zoneout regularization for all RNN layers including the attention RNN and decoder RNN. We refer to the modified encoder module as ``CBHL'' from here on.

At the decoder, encoded values are processed with an attention framework \cite{Bahdanau2014, Luong2018, DBLP:journals/corr/Graves13} to align with decoder outputs. The output from the decoder pre-net is first concatenated with the context vector from the previous time step and processed by the attention RNN. The attention mechanism relates the encoder output and output from the attention RNN by assigning probability for which input label corresponds to the output acoustic feature in the current time step. Then the context vector is computed by the weighted sum of encoder output with the probabilities the attention mechanism assigns as weights. The final output of the attention framework is the concatenation of the context vector and the output from the attention RNN. We use the LSTM cell for the attention RNN. For an attention mechanism, we use forward attention without a transition agent \cite{Zhang2018} instead of additive attention \cite{Bahdanau2014}. As mentioned in \cite{Zhang2018}, the forward attention accelerates the alignment learning speed.
In addition, forward attention is proved to give fewer alignment errors than additive and location-sensitive attention, which are used in Tacotron and Tacotron2 \cite{He2019}, and we observed the same trends in our preliminary experiments.

The output of the attention framework is processed by the decoder RNN. We use the LSTM cell for the decoder RNN. The output of the decoder RNN is projected to the mel-spectrogram with a linear layer as the final output of the network. In addition to the mel-spectrogram, Japanese Tacotron predicts stop flags \cite{DBLP:conf/iclr/PingPGAKNRM18, Shen2017} to determine when to stop the prediction loop.

We use WaveNet \cite{oord2016wavenet} to synthesize waveforms from the mel-spectrogram. WaveNet is trained with the same mel-spectrogram used for training Japanese Tacotron. We use the same conditions for the mel-spectrogram as \cite{Wang2017}: 80 bins, 50~ms frame length, and 12.5~ms frame shift.

\subsection{Self-attention Tacotron}
\label{subsec:self-attention-tacotron}

We use self-attention Tacotron proposed in \cite{DBLP:conf/icassp/YasudaWTY19}
as the third system to be investigated. The motivation to use self-attention in Tacotron is to capture long-term dependency better to produce correct pitch accent of Japanese language, which is phrase level information. It is known that by directly connecting distant states, self-attention relieves the high burden placed on LSTM to learn long-term dependencies to sequentially propagate information over long distances \cite{Lin2017}.

Figure \ref{fig:japanese-tacotron}-B shows the network structure of self-attention Tacotron, and Table~\ref{tbl:tacotron-tacotron2} summarizes the network configuration. The difference of self-attention Tacotron from the baseline Tacotron is the existence of self-attention layers in both the encoder and decoder. We used a self-attention block based on \citet{Vaswani2017}. The self-attention has multiple heads, and a residual connection is placed after the self-attention layer. As in \cite{Lin2017}, the self-attention block is placed after the LSTM layers. The encoder has two outputs: output of the bidirectional LSTM layer and output of the self-attention layer. The two encoder outputs are aligned to output acoustic features with different attention mechanisms in dual source attention \cite{BLP:journals/corr/ZophK16}. The output from the bidirectional LSTM layer is fed to forward attention \cite{Zhang2018}, and the output from the self-attention layer is fed to additive attention \cite{Bahdanau2014}. By using attention for each encoder output, we can check how the encoder outputs are used from the decoder by visualizing the attention probability distribution. In our previous study, we discovered that the forward attention captured the source-target relationship because it showed monotonic alignment, and the additive attention captured the phrase structure.
The outputs from two attention layers are concatenated and processed by the decoder RNN followed by self-attention.

\section{Experiments}
\label{sec:experiment}

Using baseline Tacotron and self-attention Tacotron with variations of encoder structure and model parameter size, we conducted two experiments to analyze several important factors for end-to-end speech synthesis \footnote{Audio samples for our experiments can be found at https://nii-yamagishilab.github.io/yasuda-csl-tacotron-audio-samples/.}. The two experiments are about Japanese and English TTS systems to answer the following questions: 
\begin{itemize}
    \item Q1) How much can increasing model parameter size improve naturalness?
    \item Q2) How much do the differences in network structures between the CBHL encoder \cite{Wang2017} and CNN-based encoder \cite{Shen2017} affect naturalness?
    \item Q3) How much can the presence of self-attention improve naturalness?
    \item Q4) How much do our sequence-to-sequence based  systems differ from comparable pipeline systems using full-context label in terms of naturalness of synthetic speech, and where do the differences possibly come from?
\end{itemize}
The English experiment uses raw texts in addition to phones, which is not possible in the Japanese experiment. This experiment is to answer the following question:
\begin{itemize}
    \item Q5) How much is character input inferior to phone input in our Tacotron systems? What can reduce the gap between character and phone input?
\end{itemize}

\subsection{Experimental conditions for Japanese systems}
\label{subsec:condition-japanese-tts}

\paragraph{\textbf{Database}} For the Japanese speech corpus, we chose the ATR Ximera corpus \cite{kawai2006ximera}, the same Japanese speech corpus used in our previous experiment \cite{DBLP:conf/icassp/YasudaWTY19}. This corpus contains 28,959 utterances or around 46.9 hours in total duration from a female speaker. We also used the same settings as in the previous research. For linguistic features, we used phonemes and accentual type labels from manually annotated labels \cite{Luong2018}. We trimmed the beginning and ending silences from the utterances, after which the duration of the corpus was reduced to 33.5 hours. We used 27,999 utterances for training, 480 for validation, and 480 for testing. All 480 samples in the test set were used for an objective evaluation, and 141 samples in the test set were used for a subjective evaluation.

\paragraph{\textbf{Tacotron configurations}}

We built eight Tacotron models for the experiment that vary in three aspects motivated by Q1-Q3 that we want to address:
\begin{itemize} 
\item Whether the model parameter size is large or small. This is implemented mainly by changing the size of hidden layers, and the details are explained later;
\item Whether the model uses the CBHL-based encoder defined in Section~\ref{subsec:japanese-tacotron} or the CNN-based encoder in the original Tacotron2 \cite{Shen2017};
\item Whether the model uses self-attention or not. Models with self-attention are based on the proposed self-attention Tacotron in Section~\ref{subsec:self-attention-tacotron}, while other models are based on the baseline Tacotron in Section~\ref{subsec:japanese-tacotron}.
\end{itemize} 
The configurations of these eight models are also listed in the three left columns of Table~\ref{tbl:japanese-alignment-error}.

Table~\ref{tbl:tacotron-tacotron2} summarizes the configurations for the large and small parameter sizes on baseline and self-attention Tacotron.
For the small parameter size, we followed configurations used for the original experiment of Tacotron \cite{Wang2017} and self-attention Tacotron \cite{DBLP:conf/icassp/YasudaWTY19}. 
Specifically, to incorporate accentual type as input to Tacotron, we used 32-dimensional accentual type embedding besides the phoneme embedding with 224 dimensions. 
The encoder pre-net on the input phoneme embeddings has two layers: one with 224 dimensions and the other with 112 dimensions. The pre-net for accentual type input has 32 and 16 dimensions for its two layers. 
The CBHL encoder has 128 units, and its output from the bidirectional RNN has 256 dimensions. The decoder pre-net has 256 and 128 dimensions for each layer. The attention and decoder RNNs have 256 dimensions each. Location features in forward attention has 5 dimensions for filter and 10 dimensions for kernel. For the self-attention layer in self-attention Tacotron, we used 32 and 256 dimensions for the encoder and the decoder, respectively.

For the large parameter size, we followed configurations of Tacotron2 \cite{Shen2017}. We used 64 and 448 dimensions for accentual type embedding and phoneme embedding, respectively. 
The encoder pre-net on phoneme increases the sizes of two layers to 448 and 224 dimensions, respectively. The layer sizes in pre-net on accentual type are increased to 64 and 32 dimensions. 
The CBHL encoder has 256 units, and its output from the bidirectional RNN is 512 dimensions. The decoder pre-net has 256 dimensions for each layer. The attention and decoder RNN have 128 and 1024 dimensions, respectively. For the self-attention layer in self-attention Tacotron, we used 64 and 1024 dimensions for the encoder and decoder, respectively. These parameter configurations are also shown in figure \ref{fig:japanese-tacotron}-A for baseline Tacotron and figure \ref{fig:japanese-tacotron}-B for self-attention Tacotron.


For models using the CNN-based encoder, we set the hidden layers of the encoder to 256 and 512 dimensions for the small and large parameter sizes, respectively. 

For all the eight experimental models, we further added CNN-based post-net of Tacotron2 for a fair comparison. Examples of post-net for large baseline and self-attention Tacotron are plotted in figure~\ref{fig:japanese-tacotron}.

\paragraph{\textbf{Tacotron and Wavenet training}}
We used L1 loss for the mel-spectrogram, and binary cross entropy loss for the stop flag. The models were optimized with an Adam optimizer \cite{DBLP:journals/corr/KingmaB14}, with exponential learning rate decay from $10^{-4}$ to $10^{-5}$. We applied L2 regularization with a weight of $10^{-6}$ for baseline Tacotron and $10^{-7}$ for self-attention Tacotron with large parameter size.  To avoid increasing training time, we used a reduction factor of two, the same value as in the previous work \cite{DBLP:conf/icassp/YasudaWTY19}, instead of the reduction factor of one used in Tacotron2 \cite{Shen2017}. We did not enable dropout during inference unlike Tacotron2 \cite{Shen2017}. For waveform synthesis, we used the same $\mu$-law WaveNet \cite{oord2016wavenet} model from the previous research, which was trained with ground truth mel-spectrograms of the same condition as acoustic features used by our Tacotron systems: 50~ms frame length and 12.5~ms frame shift. 


\paragraph{\textbf{Objective evaluation}}
We measured the alignment error rate on the eight experimental Japanese Tacotron systems. There are three types of fatal alignment errors (discontinuous alignment, incomplete alignment, and overestimated duration) that can be easily detected with software or human eyes by examining the attention probability distribution. Specifically, the discontinuous alignment error is indicated by a skip or repeat of a transcription and can be detected by finding non-monotonic jumps of the attention distribution mode. The incomplete alignment error is indicated by a premature termination of a transcription and can be detected by finding an alignment distribution mode that does not reach the final input position at the final time step. The overestimated duration error is signified by an excessively prolonged phoneme and can be detected by finding an attention distribution mode staying in the same position for a long time. 

We measured alignment error rates on each system several times. Each time, we initialized the system with a different random seed and trained it from scratch. After generating the 480 test set utterances, we counted the alignment errors using in-house software.


\paragraph{\textbf{Subjective evaluation}} We conducted a listening test about naturalness of synthetic speech. We recruited 213 native Japanese listeners via crowdsourcing. 
Listeners were asked to evaluate 28 samples in one set, using a five-grade mean opinion score (MOS) metric.
One listener could evaluate at most ten sets. We collected 19,880 data points in total. We checked the statistical significance of the scores between systems with a Mann-Whitney rank test \cite{Mann-WhitneyRankTest}.

The listening test contained 14 systems including natural samples, three analysis-by-synthesis systems\footnote{Analysis-by-synthesis systems are the reference systems using neural vocoders given natural acoustic features extracted from the test set utterances for waveform generation. They simulate TTS systems that can perfectly convert input text into acoustic features.}, two pipeline TTS systems, and the eight Tacotron systems. The three analysis-by-synthesis systems were WaveNet models used for the DNN based SPSS system using the vocoder parameter, for the vocoder-free TTS system using mel-spectrogram with 5~ms frame shift, and for the Tacotron systems using the mel-spectrogram with 12.5~ms frame shift. The two pipeline systems were the DNN based SPSS and vocoder-free TTS systems using an autoregressive acoustic model as described earlier. These pipeline systems used manually annotated full-context labels \cite{Luong2018}, of which the subsets are the phoneme and accentual type labels used for all the Tacotron systems. With these pipeline systems, the experiment is expected to answer Q4 posed at the beginning of this section.

Since we trained the eight Tacotron models multiple times during objective evaluation, we selected the instance with the lowest alignment error rate and used the corresponding generated utterances for the listening test. 

\subsection{Results of Japanese systems}
\label{subsec:result-japanese-tts}

\begin{table}[t]
\caption{
Alignment error rate on Japanese test set. Each system was evaluated three times, and each time it was trained from scratch with a different initial random seed. Ave. indicates average error rate. Values in bold correspond to those evaluated in a listening test.
}
\label{tbl:japanese-alignment-error}
\small{
\setlength{\tabcolsep}{4pt}
\begin{center}
\begin{tabular}{|c|c|c||c|cccc|}\hline
Para. & \multirow{2}{*}{Encoder} & Self- & \# Para. & \multicolumn{4}{c|}{Alignment error rate (\%)}\\
size & ~ & attention &  ($1\times{10}^6$) & Ave. & 1 & 2 & 3\\\hline\hline
\multirow{4}{*}{Small}  & \multirow{2}{*}{CBHL}     & -  & 11.3 & 2.4 & \textbf{0.4} & 2.7 & 4.2 \\\cline{3-8}
~            &  & \checkmark & 11.6 & 11.5 & \textbf{6.0} & 7.9 & 20.6\\\cline{2-8}
~            & \multirow{2}{*}{CNN}              & - & 9.2 & 2.4 & \textbf{0.6} & 1.7 & 5.0\\\cline{3-8}
~            &               & \checkmark & 9.6 & 18.2 & \textbf{4.8} & 14.2 & 35.6\\\hline
\multirow{4}{*}{Large}  & \multirow{2}{*}{CBHL} & - & 35.8 & 0.2 & \textbf{0.2} & 0.2 & 0.2 \\\cline{3-8}
~            			   & ~              & \checkmark & 41.6 & 0.3 & \textbf{0.2} & 0.2 & 0.4 \\\cline{2-8}
~            & \multirow{2}{*}{CNN} & - & 27.2 & 0.2 & \textbf{0.2} & 0.2 & 0.2\\\cline{3-8}
~            & ~              & \checkmark & 32.8 & 0.2 & \textbf{0.2} & 0.2 & 0.2\\\hline
\end{tabular}
\end{center}
}
\end{table}

Table \ref{tbl:japanese-alignment-error} shows the alignment error rates of each system. 
From this table, we can first see that all systems with the large parameter size showed low alignment error rates and were stable over different runs. Increasing parameter size improved and stabilized the learning of alignments. Second, 
systems with self-attention and small parameter size made more alignment errors and were quite unstable across different runs. Through further investigation, we found that a combination of post-net and self-attention made the alignments unstable in the small-parameter-size configurations. In general, when systems have small parameter sizes, their learning of the alignment is sensitive to the initial parameters and network structures.



\begin{figure}[t]
	\centering
		\includegraphics[width=1\columnwidth]{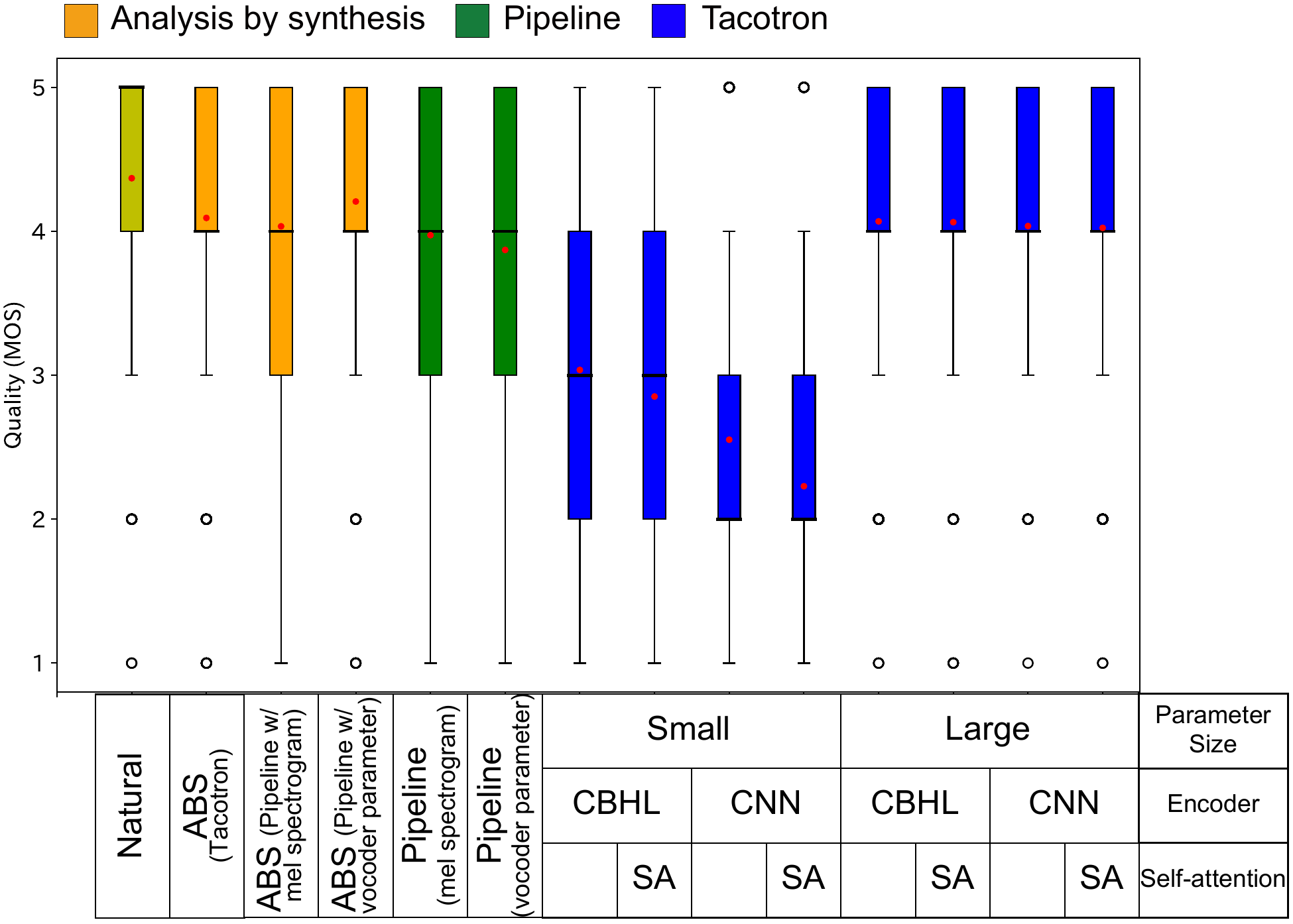}
	\caption{Box plot for a results of Japanese listening test. Red dots indicate mean, and black bars indicate median.}
	\label{fig:japanese-listening-test-big}
\end{figure}

\begin{table*}[t]
\caption{Mann-Whitney rank test for Japanese listening test. Cell in \textcolor{tsigc}{this color denotes statistical significance ($p \le 0.01$)} while \textcolor{tnsigc}{this color denotes $p > 0.01$}.}
\vspace{-7px}
\footnotesize
\label{tbl:japanese-u-test}
\begin{center}
\setlength{\tabcolsep}{4pt}
\begin{tabular}{|c|c|c|c||p{0.5cm}|p{0.5cm}|p{0.5cm}|p{0.5cm}|p{0.5cm}|p{0.5cm}||p{0.5cm}|p{0.5cm}|p{0.5cm}|p{0.5cm}|p{0.5cm}|p{0.5cm}|p{0.5cm}|p{0.5cm}|c}
\hhline{*{18}{-}}
\multicolumn{4}{|c||}{~} & \multirow{4}{*}{\rotatebox{90}{Natural}} & \multicolumn{3}{c|}{ABS} & \multicolumn{2}{c||}{Pipeline} & \multicolumn{8}{c|}{Tacotron} & \\ \hhline{|~~~~||~|-----||---------}
\multicolumn{4}{|c||}{~} & ~ & \multirow{3}{*}{\rotatebox{90}{\shortstack{Tacotron \\ Mel.}}} & \multirow{3}{*}{\rotatebox{90}{\shortstack{Pipeline \\ Mel.}}} & \multirow{3}{*}{\rotatebox{90}{\shortstack{Vocoder \\ Para.}}} & \multirow{3}{*}{\rotatebox{90}{\shortstack{Pipeline \\ Mel.}}} & \multirow{3}{*}{\rotatebox{90}{\shortstack{Vocoder \\ Para.}}} & \multicolumn{4}{c|}{Small} & \multicolumn{4}{c|}{Large} & \multicolumn{1}{c|}{Para. size} \\ \hhline{|~~~~||~|~|~|~|~|~||--------|-|}
\multicolumn{4}{|c||}{~} & ~ &  &  & & & & \multicolumn{2}{c|}{CBHL} & \multicolumn{2}{c|}{CNN} & \multicolumn{2}{c|}{CBHL} & \multicolumn{2}{c|}{CNN} & \multicolumn{1}{c|}{Encoder}    \\ \hhline{|~~~~||~|~|~|~|~|~||--------|-|}
\multicolumn{4}{|c||}{~} & ~ &  &  &  & & & & \checkmark & ~ & \checkmark & ~ & \checkmark & ~ & \checkmark & \multicolumn{1}{c|}{Self-att.} \\ \hhline{|====||=|=|=|=|=|=||=|=|=|=|=|=|=|=|=|} 

\multicolumn{4}{|c||}{{Natural}} 							    & ~ & \tsig & \tsig & \tsig & \tsig & \tsig & \tsig & \tsig & \tsig & \tsig & \tsig & \tsig & \tsig & \tsig &  \\ \hhline{|----||-|-|-|-|-|-||-|-|-|-|-|-|-|-|~}
\multirow{3}{*}{{ABS}} & \multicolumn{3}{c||}{Tacotron Mel. spec.} & \tsig & ~ & \tnsig & \tsig & \tsig & \tsig & \tsig & \tsig & \tsig & \tsig & \tnsig & \tnsig & \tsig & \tsig & ~\\ 
\hhline{|~|---||-|-|-|-|-|-||-|-|-|-|-|-|-|-|~}
~ 				  & \multicolumn{3}{c||}{Pipeline Mel. spec.}  & \tsig & \tnsig & ~ & \tsig & \tsig & \tsig & \tsig & \tsig & \tsig & \tsig & \tnsig & \tnsig & \tnsig & \tnsig & ~\\ 
\hhline{|~|---||-|-|-|-|-|-||-|-|-|-|-|-|-|-|~}
~ 				  & \multicolumn{3}{c||}{Vocoder para.}		& \tsig & \tsig & \tsig & ~ & \tsig & \tsig & \tsig & \tsig & \tsig & \tsig & \tsig & \tsig & \tsig & \tsig & ~\\ 
\hhline{|-|---||-|-|-|-|-|-||-|-|-|-|-|-|-|-|~}
\multirow{2}{*}{{Pipline}} & \multicolumn{3}{c||}{Pipeline Mel. spec.}  		& \tsig & \tsig & \tsig & \tsig & ~ & \tnsig & \tsig & \tsig & \tsig & \tsig & \tsig & \tsig & \tnsig & \tnsig & ~\\ 
\hhline{|~|---||-|-|-|-|-|-||-|-|-|-|-|-|-|-|~}
~ 				     & \multicolumn{3}{c||}{Vocoder para.}   	& \tsig & \tsig & \tsig & \tsig & \tnsig & ~ & \tsig & \tsig & \tsig & \tsig & \tsig & \tsig & \tsig & \tsig & ~ \\ 
\hhline{|=|===||=|=|=|=|=|=||=|=|=|=|=|=|=|=|~} 
\multirow{8}{*}{\rotatebox{90}{Tacotron}} & \multirow{4}{*}{{Small}} 	& \multirow{2}{*}{{CBHL}} & - & \tsig & \tsig & \tsig & \tsig & \tsig & \tsig & ~ & \tsig & \tsig & \tsig & \tsig & \tsig & \tsig & \tsig ~\\ 
\hhline{|~|~|~|-||-|-|-|-|-|-||-|-|-|-|-|-|-|-|~} 
~ 					& 					& 					& \checkmark	& \tsig & \tsig & \tsig & \tsig & \tsig & \tsig & \tsig & ~ & \tsig & \tsig & \tsig & \tsig & \tsig & \tsig & ~ \\ 
\hhline{|~|~|-|-||-|-|-|-|-|-||-|-|-|-|-|-|-|-|~}
~ 					& 					& \multirow{2}{*}{{CNN}}	& -		 	& \tsig & \tsig & \tsig & \tsig & \tsig & \tsig & \tsig & \tsig & ~ & \tsig & \tsig & \tsig & \tsig & \tsig & ~\\ 
\hhline{|~|~|~|-||-|-|-|-|-|-||-|-|-|-|-|-|-|-|~} 
~ 					& 					& ~ 					& \checkmark 	& \tsig & \tsig & \tsig & \tsig & \tsig & \tsig & \tsig & \tsig & \tsig & ~ & \tsig & \tsig & \tsig & \tsig & ~\\ 
\hhline{|~|-|-|-||-|-|-|-|-|-||-|-|-|-|-|-|-|-|~}  
~ 					& \multirow{4}{*}{{Large}} 	& \multirow{2}{*}{{CBHL}} 	& - 			& \tsig & \tnsig & \tnsig & \tsig & \tsig & \tsig & \tsig & \tsig & \tsig & \tsig & ~ & \tnsig & \tnsig & \tnsig & ~\\ 
\hhline{|~|~|~|-||-|-|-|-|-|-||-|-|-|-|-|-|-|-|~}  
~ 					& 					& ~ 					& \checkmark 	& \tsig & \tnsig & \tnsig & \tsig & \tsig & \tsig & \tsig & \tsig & \tsig & \tsig & \tnsig & ~ & \tnsig & \tnsig & ~\\ 
\hhline{|~|~|-|-||-|-|-|-|-|-||-|-|-|-|-|-|-|-|~}  
~ 					& 					& \multirow{2}{*}{{CNN}} 	& -			& \tsig & \tsig & \tnsig & \tsig & \tnsig & \tsig & \tsig & \tsig & \tsig & \tsig & \tnsig & \tnsig & ~ & \tnsig & ~\\ 
\hhline{|~|~|~|-||-|-|-|-|-|-||-|-|-|-|-|-|-|-|~} 
~ 					& 					& ~ 					& \checkmark 	& \tsig & \tsig & \tnsig & \tsig & \tnsig & \tsig & \tsig & \tsig & \tsig & \tsig & \tnsig & \tnsig & \tnsig & ~ & ~\\ \hhline{|-|-|-|-||-|-|-|-|-|-||-|-|-|-|-|-|-|-|~}  
\end{tabular}
\normalsize
\end{center}
\end{table*}

Figure \ref{fig:japanese-listening-test-big} shows the results of the listening test in Japanese, and Table \ref{tbl:japanese-u-test} lists the outcome from the statistical significance test. We can see a significant gap between Tacotron systems with small and large parameter sizes. All systems with the small parameter size had low scores: baseline Tacotron had scores of $3.04\pm 0.04$ and $2.55\pm 0.04$ for the CBHL and CNN encoders, respectively, and self-attention Tacotron systems had $2.85\pm 0.04$ and $2.23\pm 0.04$ for the CBHL and CNN encoders, respectively. For both baseline Tacotron and self-attention Tacotron, the CBHL encoder performed better than the CNN encoder in the small parameter size conditions. On the other hand, Tacotron systems with the large parameter size had high MOSs of about 4.0.
We listened to the samples that had average MOSs of less than 2.5 from the small-parameter-size systems and found incorrect pitch accents in them. Furthermore, samples from systems using the CNN encoder with the small parameter size generally sounded flatter in pitch than corresponding systems using the CBHL encoder, which is probably why listeners rated them negatively.
Some lowly rated samples from small self-attention Tacotron also contained fatal alignment errors, which were identified by the alignment error detector. 

The difference among Tacotron systems with the large parameter size is not statistically significant as  Table~\ref{tbl:japanese-u-test} shows, so the presence of self-attention and difference in encoder network structure did not significantly affect the naturalness of synthetic speech. Furthermore, these Tacotron systems had slightly higher scores than the pipeline system using the mel-spectrogram, which had an MOS of $3.97 \pm 0.04$. The Tacotron systems using the CBHL encoder had statistically significant differences from the pipeline system using mel-spectrogram, but the Tacotron systems using the CNN based encoder did not. The pipeline system using the vocoder parameter had a lower MOS ($3.87 \pm 0.04$) than the pipeline system using the mel-spectrogram and the Tacotron system with the large parameter size.

\subsection{Discussion on results of Japanese experiment}

This experiment showed: Q1) Increasing parameter size greatly improved naturalness. Q2) The CBHL encoder performed better than the CNN encoder for systems with the small parameter size. The difference in encoder structure however did not make much difference for systems with the large parameter size. Q3) Self-attention did not improve naturalness for models with the large parameter size. Self-attention was expected to improve naturalness for models with the small parameter size in accordance with our previous research \cite{DBLP:conf/icassp/YasudaWTY19}, but any improvement was undercut by alignment errors caused by poor training in this experiment. Q4) The Tacotron system with the large parameter size using the CBHL encoder slightly outperformed the pipeline system using the same mel-spectrogram and the same neural vocoder.  

In the previous research \cite{DBLP:conf/icassp/YasudaWTY19}, our Tacotron systems did not match the pipeline systems in terms of the quality of synthetic speech, and we hypothesized that one possible reason was the linguistic feature limitation, but this hypothesis turns out to be incorrect. The main reason is simply the parameter size.

When a model has sufficient parameter capacity, we confirmed that the Tacotron systems can learn to compensate for the lack of other rich information and accurately predict mel-spectrogram without additional input contextual labels such as part-of-speech tags and syntactic information that are used in the pipeline systems.
The score of the best Tacotron system (large parameter size and CBHL encoder) in our experiment scored higher than that of the pipeline system, but, as expected, the difference is smaller than that reported in the literature since both systems use autoregressive modeling and the same neural vocoder. 

There are many differences between Tacotron and Tacotron2 apart from the use of neural vocoders as we described earlier. Our result, however, reveals that such changes are not fundamental differences and do not bring about significant improvements, at least, for our Japanese TTS systems using phonemes. Our analysis results indicate that the major improvements simply come from the model capacity. 

However, we also observed that the linguistic feature limitation affected accuracy of pitch accent prediction for a few systems. More specifically, we found incorrect pitch accent from audio samples generated using the systems with the small parameter size. For pitch accent information, we fed accentual type labels, which indicate the accentual phrase boundary and type of pitch accent but do not explicitly indicate where to drop pitch. 
Therefore, we think that the systems with the small parameter size were occasionally unable to resolve the position to drop pitch given accentual type labels alone. This is consistent with results reported in \citet{Fujimoto2019}, where an abstract pitch contour pattern is used such as high and low labels plus an accentual phrase boundary.

\subsection{Experimental conditions for English systems}

\paragraph{\textbf{Database}} We chose the Blizzard 2011 corpus \cite{Blizzard2011} as a speech corpus to construct English TTS models. This corpus contains 12,092 utterances or around 17 hours in total duration from a female speaker. This time we used characters or phones for input. The Blizzard 2011 corpus is suitable to compare character and phone input because it has a wider variety of unique word types (18,695 unique words) \cite{Taylor2019}, so we can thoroughly check the generalizability of our systems using character input. We obtained phone labels by using Flite \cite{Black2001Flite}. All characters were converted into lowercase, but no elaborate text normalization was performed. We trimmed the beginning and end silences from the utterances, after which the duration of the corpus was reduced to 15.7 hours. We used 11,092 utterances for training, 500 for validation, and 500 for testing.

\paragraph{\textbf{Tacotron configurations}} 
Similar to the Japanese experiment, we built 16 English Tacotron models in order to answer the questions (Q1-3 and Q5) posed at the beginning of this section for English: each model is either small or large in model parameter size for Q1, uses CBHL or CNN-based encoder for Q2, has or does not have self-attention for Q3, and takes phone or raw text as input for Q5. Different from Japanese models, however, we removed the accentual type embedding from the input because it is unnecessary for English TTS. 
Accordingly, we allocated 256 dimensions for the small-parameter-size condition and 512 dimensions for the large-parameter-size condition to character or phone embedding. Two layers in encoder pre-net were configured to have 256 and 128 dimensions for the small-parameter-size condition and 512 and 256 dimensions for the large-parameter-size condition. 

\paragraph{\textbf{Objective evaluation}}
We used the alignment error rates as objective metrics in the same way as the Japanese experiment did in Section \ref{subsec:condition-japanese-tts}.
We measured the error rates of the 16 English Tacotron models on the test set with 500 utterances.

\paragraph{\textbf{Subjective evaluation}} We conducted a listening test about naturalness of synthetic speech similarly to the Japanese test.
We recruited 49 native English listeners via crowdsourcing. Listeners were asked to evaluate 36 samples using the five-grade MOS metric in one set. One listener could evaluate at most 20 sets. We collected 17,928 evaluations in total. 

For the subjective evaluation, we excluded Tacotron systems with the small parameter size since they have obviously worse objective evaluation results as described later and may cause the ceiling effect that disturbs the analysis of other factors. Therefore, the listening test included 12 systems in total: eight Tacotron systems, one DNN based SPSS system using vocoder parameter \cite{Luong2018}, two analysis-by-synthesis systems from WaveNet used for the pipeline and Tacotron, and natural samples.

\subsection{Results of English systems}

Tables \ref{tbl:english-alignment-error} and \ref{tbl:english-alignment-error-character} shows the number of samples containing fatal alignment errors from systems using phones and characters as an input, respectively.
All systems with the small model parameter size showed a relatively large alignment error rate, ranging from 7.6~\% to 23.4~\% on average. Among the systems with the small parameter size, systems using self-attention had especially higher alignment error rates. One possible reason is that the combination of post-net and self-attention made alignment unstable as we described in section \ref{subsec:result-japanese-tts}. All systems with the large model parameter size had only small alignment errors. The alignment error rates from these systems were consistent across multiple runs with different random initial seeds. 

Overall, the large-parameter-size configuration helped to improve and stabilize alignment and its learning, whereas the small-parameter-size configuration made it unstable and sensitive to initial parameters and network structure. These results were consistent with those from the Japanese experiment.



\begin{table}[t]
\caption{
Alignment error rate on English test set. Each system was evaluated for three times, and each time it was trained from scratch with a different initial random seed. Ave. indicates average error rate. Values in bold correspond to those evaluated in a listening test.
}

\small{
\setlength{\tabcolsep}{4pt}
\begin{center}
\subfigure[Systems using phone as input]{
\label{tbl:english-alignment-error}
\begin{tabular}{|c|c|c||c|cccc|}\hline
Para.& \multirow{2}{*}{Encoder} & {Self-} &  \# Para. &  \multicolumn{4}{c|}{Alignment error rate (\%)}\\
size & ~ & {attention} & ($1\times{10}^6$) & Ave. & 1 & 2 & 3\\\hline\hline
\multirow{4}{*}{Small}   & \multirow{2}{*}{CBHL}    & -             & 7.9 & 7.6 & 6.2 & 8.6 & 9.0\\\cline{3-8}
~                        &                          & \checkmark    & 12.1 & 17.9 & 10.0 & 10.4 & 33.2\\\cline{2-8}
~                        & \multirow{2}{*}{CNN}     & -             & 9.2  & 9.3 & 5.0 & 5.2 & 17.8\\\cline{3-8}
~                        & ~                        & \checkmark    & 9.9  & 15.6 & 13.0 & 16.6 & 17.2\\\hline
\multirow{4}{*}{Large}   & \multirow{2}{*}{CBHL}    & -             & 36.7 & 1.0 & \textbf{0.6} & 0.8 & 1.6\\\cline{3-8}
~                        & ~                        & \checkmark    & 47.6 & 0.6 & \textbf{0.2} & 0.6 & 1.0\\\cline{2-8}
~                        & \multirow{2}{*}{CNN}     & -             & 27.2 & 1.1 & \textbf{0.2} & 1.4 & 1.6\\\cline{3-8}
~                        & ~                        & \checkmark    & 38.1 & 1.0 & 0.8 & 1.0 & \textbf{1.2}\\\hline
\end{tabular}

}
\subfigure[Systems using character as input]{
\label{tbl:english-alignment-error-character}
\begin{tabular}{|c|c|c||c|cccc|}\hline
Para.& \multirow{2}{*}{Encoder} & {Self-} &  \# Para. &  \multicolumn{4}{c|}{Alignment error rate (\%)}\\
size & ~ & {attention} & ($1\times{10}^6$) & Ave. & 1 & 2 & 3\\\hline\hline
\multirow{4}{*}{Small}   & \multirow{2}{*}{CBHL}    & -             & 11.3 & 14.9 & 7.4 & 8.6 & 28.8\\\cline{3-8}                        &                          & \checkmark    & 12.1 & 23.4 & 18.8 & 21.8 & 29.6\\\cline{2-8}
~                        & \multirow{2}{*}{CNN}     & -             & 9.2  & 11.1 & 6.6 & 10.8 & 15.8\\\cline{3-8}
~                        & ~                        & \checkmark    & 9.9  & 18.0 & 15.0 & 16.8 & 22.2\\\hline
\multirow{4}{*}{Large}   & \multirow{2}{*}{CBHL}    & -             & 36.7 & 1.1 & 0.8 & \textbf{1.0} & 1.6\\\cline{3-8}
~                        & ~                        & \checkmark    & 47.6 & 0.8 & 0.4 & \textbf{1.0} & 1.0\\\cline{2-8}
~                        & \multirow{2}{*}{CNN}     & -             & 27.2 & 0.5 & \textbf{0.2} & 0.6 & 0.6\\\cline{3-8}
~                        & ~                        & \checkmark    & 38.1 & 0.7 & \textbf{0.4} & 0.6 & 1.0\\\hline
\end{tabular}
}

\end{center}
}
\end{table}

\begin{figure}[t]
	\centering
		\includegraphics[width=1\columnwidth]{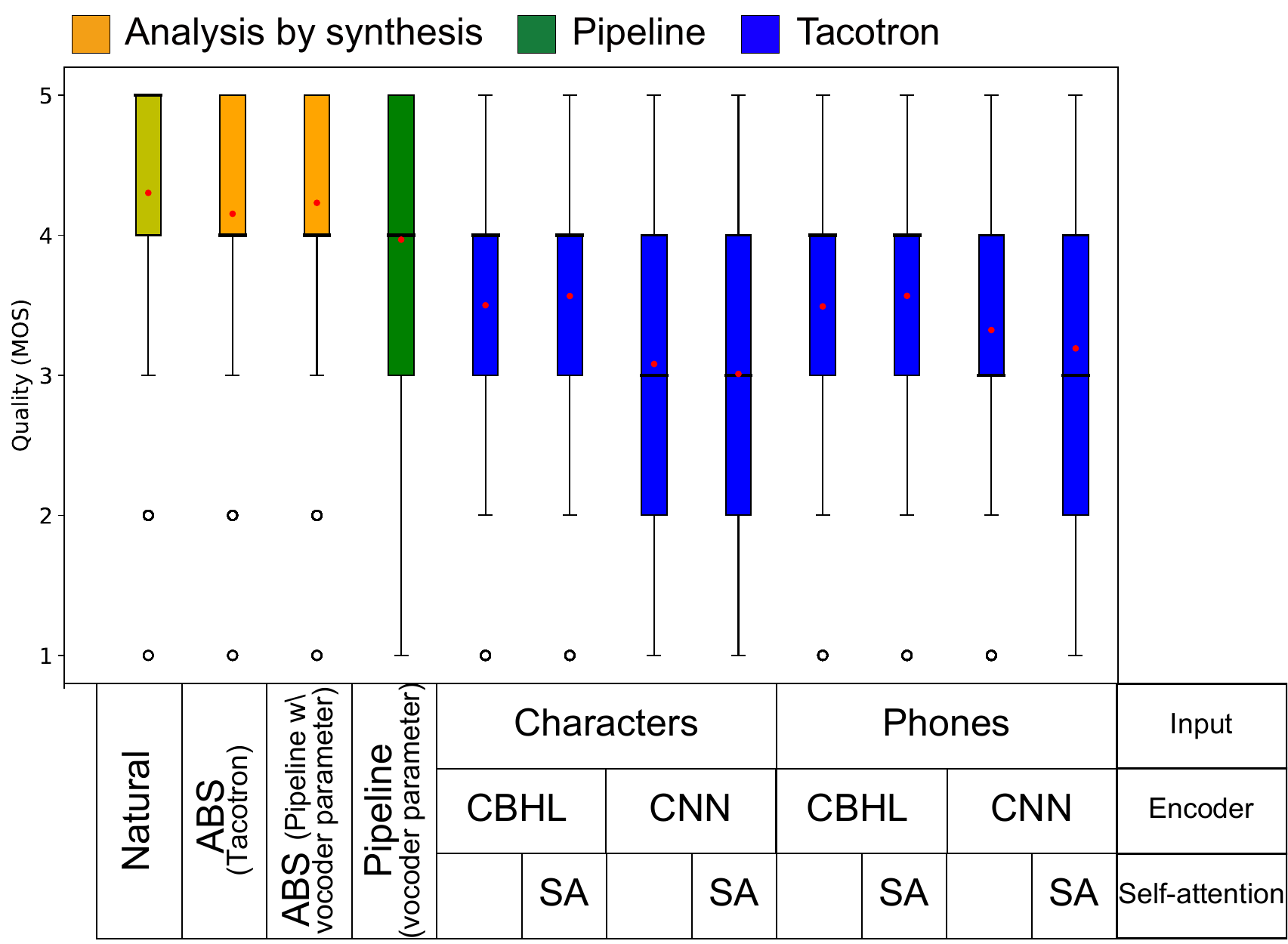}
	\caption{Box plot for the results of English listening test. Red dots indicate mean, and black bars indicate median.}
	\label{fig:english-listening-test}
\end{figure}

As we described earlier, we conducted a listening test to investigate the impacts of configurations other than the parameter size by using systems with the large parameter size.
Figure \ref{fig:english-listening-test} shows the results of the listening test in English. Refer to Table \ref{tbl:english-u-test} for the statistical significance of the test. First, we can see that when the encoder was based on CNN, both systems with and without self-attention using phones had higher scores than corresponding systems using characters. The baseline Tacotron with the CNN based encoder had $3.08 \pm 0.05$ when characters were used and $3.32 \pm 0.05$ when phones were used. Self-attention Tacotron with the CNN based encoder had $3.01 \pm 0.05$ when characters were used and $3.19 \pm 0.05$ when phones were used. 

Second, interestingly, we can see that when the CBHL encoder was used, the score gap caused by the difference in input features vanished. The baseline Tacotron with the CBHL encoder had $3.50 \pm 0.04$ when characters were used and $3.49 \pm 0.04$ when phones were used. Self-attention Tacotron with the CBHL encoder had $3.57 \pm 0.04$ both when characters and phones were used. 

Third, self-attention did not show statistical significance except for systems using the CNN encoder and phone inputs, and their score difference seemed to represent just their alignment error rates.  
Finally, the Tacotron systems did not match the pipeline system in terms of naturalness.

\begin{table*}[t]
\caption{Mann-Whitney rank test for English listening test. Cell in \textcolor{tsigc}{this color denotes statistical significance ($p \le 0.01$)} while \textcolor{tnsigc}{this color denotes $p > 0.01$}.}
\vspace{-7px}
\footnotesize
\label{tbl:english-u-test}
\begin{center}
\setlength{\tabcolsep}{4pt}
\begin{tabular}{|c|c|c|c||p{0.5cm}|p{0.5cm}|p{0.5cm}|p{0.5cm}||p{0.5cm}|p{0.5cm}|p{0.5cm}|p{0.5cm}|p{0.5cm}|p{0.5cm}|p{0.5cm}|p{0.5cm}|c}
\hhline{*{16}{-}}
\multicolumn{4}{|c||}{~} & \multirow{4}{*}{\rotatebox{90}{Natural}} & \multicolumn{2}{c|}{ABS} & \multirow{4}{*}{\rotatebox{90}{\shortstack{Pipeline \\ Vocoder}}} & \multicolumn{8}{c|}{Tacotron (Large  parameter size)} & \\ \hhline{|~~~~||~|--~||---------}
\multicolumn{4}{|c||}{~} & ~ & \multirow{3}{*}{\rotatebox{90}{\shortstack{Tacotron \\ Mel.}}} & \multirow{3}{*}{\rotatebox{90}{\shortstack{Vocoder \\ Para.}}} &  & \multicolumn{4}{c|}{Charater} & \multicolumn{4}{c|}{Phone} & \multicolumn{1}{c|}{Input} \\ \hhline{|~~~~||~|~|~|~||--------|-|}
\multicolumn{4}{|c||}{~} & ~ &  &  & & \multicolumn{2}{c|}{CBHL} & \multicolumn{2}{c|}{CNN} & \multicolumn{2}{c|}{CBHL} & \multicolumn{2}{c|}{CNN} & \multicolumn{1}{c|}{Encoder}    \\ \hhline{|~~~~||~|~|~|~||--------|-|}
\multicolumn{4}{|c||}{~} & ~ &  &  &  & & \checkmark & ~ & \checkmark & ~ & \checkmark & ~ & \checkmark & \multicolumn{1}{c|}{Self-att.} \\ \hhline{|====||=|=|=|=||=|=|=|=|=|=|=|=|=|} 

\multicolumn{4}{|c||}{{Natural}} 							    & ~ &  \tsig & \tsig & \tsig & \tsig & \tsig & \tsig & \tsig & \tsig & \tsig & \tsig & \tsig &  \\ \hhline{|----||-|-|-|-||-|-|-|-|-|-|-|-|~}
\multirow{2}{*}{{ABS}} & \multicolumn{3}{c||}{Tacotron Mel. spec.} & \tsig & ~ & \tsig & \tsig & \tsig & \tsig & \tsig & \tsig & \tsig & \tsig & \tsig & \tsig & ~\\ 
\hhline{|~|---||-|-|-|-||-|-|-|-|-|-|-|-|~}
~ 				  & \multicolumn{3}{c||}{Vocoder para.}		& \tsig & \tsig & ~ & \tsig &  \tsig & \tsig & \tsig & \tsig & \tsig & \tsig & \tsig & \tsig & ~\\ 
\hhline{|-|---||-|-|-|-||-|-|-|-|-|-|-|-|~}
\multicolumn{4}{|c||}{Pipeline vocoder para.}   	& \tsig & \tsig & \tsig & ~ & \tsig & \tsig & \tsig &  \tsig & \tsig & \tsig & \tsig & \tsig & ~ \\ 
\hhline{|=|===||=|=|=|=||=|=|=|=|=|=|=|=|~} 
\multirow{8}{*}{\rotatebox{90}{\shortstack{Tacotron \\ (Large parameter size)}}} & \multirow{4}{*}{{Character}} 	& \multirow{2}{*}{{CBHL}} & - & \tsig & \tsig & \tsig & \tsig & ~ & \tnsig & \tsig & \tsig & \tnsig & \tnsig & \tsig & \tsig & ~\\ 
\hhline{|~|~|~|-||-|-|-|-||-|-|-|-|-|-|-|-|~} 
~ 					& 					& 					& \checkmark	& \tsig & \tsig & \tsig & \tsig & \tnsig ~ & & \tsig & \tsig & \tnsig & \tnsig & \tsig & \tsig & ~\\  
\hhline{|~|~|-|-||-|-|-|-||-|-|-|-|-|-|-|-|~}
~ 					& 					& \multirow{2}{*}{{CNN}}	& -		 	& \tsig & \tsig & \tsig & \tsig & \tsig & \tsig & ~ & ~\tnsig & \tsig & \tsig & \tsig & \tsig & ~\\ 
\hhline{|~|~|~|-||-|-|-|-||-|-|-|-|-|-|-|-|~} 
~ 					& 					& ~ 					& \checkmark 	& \tsig & \tsig & \tsig & \tsig & \tsig & \tsig & \tnsig & ~ & \tsig & \tsig & \tsig & \tsig & ~\\ 
\hhline{|~|-|-|-||-|-|-|-||-|-|-|-|-|-|-|-|~}  
~ 					& \multirow{4}{*}{{Phone}} 	& \multirow{2}{*}{{CBHL}} 	& - 			& \tsig &  \tsig & \tsig & \tsig & \tnsig & \tnsig & \tsig & \tsig & ~ & \tnsig & \tsig & \tsig & ~\\ 
\hhline{|~|~|~|-||-|-|-|-||-|-|-|-|-|-|-|-|~}  
~ 					& 					& ~ 					& \checkmark 	& \tsig & \tsig &  \tsig & \tsig & \tnsig & \tnsig & \tsig & \tsig & \tnsig & ~ & \tsig& \tsig & ~\\ 
\hhline{|~|~|-|-||-|-|-|-||-|-|-|-|-|-|-|-|~}  
~ 					& 					& \multirow{2}{*}{{CNN}} 	& -			& \tsig & \tsig & \tsig & \tsig & \tsig & \tsig & \tsig & \tsig & \tsig & \tsig & ~ & \tsig & ~\\ 
\hhline{|~|~|~|-||-|-|-|-||-|-|-|-|-|-|-|-|~} 
~ 					& 					& ~ 					& \checkmark 	& \tsig & \tsig & \tsig & \tsig & \tsig & \tsig & \tsig & \tsig & \tsig & \tsig & \tsig &  & ~\\ \hhline{|-|-|-|-||-|-|-|-||-|-|-|-|-|-|-|-|~}  
\end{tabular}
\normalsize
\end{center}
\end{table*}

\subsection{Discussion on results of English experiment}

This experiment showed: Q1) Increasing model parameter size helps to learn and predict alignments between the source and target more robustly. Q2\&5) When the CNN-based encoder was used, models using character input were inferior to models using phone input. However, when the CBHL encoder was used, models using character input could perform as well as models using phone input.
Q3) Self-attention did not affect naturalness. 
Q4) Our Tacotron systems were outperformed by the pipeline system.

We think that insights obtained from Q2 and Q5 are interesting. According to our results, the simplification of encoder architecture adopted for Tacotron2 does not improve the performance. Instead, it worsens the performance when the inputs are characters.  Pronunciation cannot always be accurately inferred from characters in English, whereas phones enable pronunciation ambiguities to be avoided. Therefore, it is reasonable that systems using phones as inputs scored higher than systems using characters, and this is confirmed by other sequence-to-sequence based TTS research \cite{DBLP:conf/iclr/PingPGAKNRM18, Zhang2019}. A reasonable conclusion that phones work better than characters was confirmed in an experiment using large-scale corpora in three languages including English \cite{Zhang2019}. However, their system used the CNN based encoder from Tacotron2 \cite{Zhang2019}, so their conclusion may not hold for other encoder structures such as the CBHL encoder.

When the CBHL encoder was used, systems using characters produced the same quality as systems using phones as input. This suggests that the particular network structure in the CBHL encoder can learn to disambiguate underlying phones given characters. A similar result was reported by \citet{Yang2019} in that the CBHL encoder worked better than a CNN plus self-attention encoder for a transformer-like sequence-to-sequence system using character input in the same corpus. The CBHL encoder was originated from a word embedding method that relies on only character-level input \cite{DBLP:conf/aaai/KimJSR16}. They compared nearest-neighbor words on the basis of cosine distance of word vectors obtained before and after highway layers and found that the highway layers could encode semantic features that were not discernible from orthography, whereas representations before highway layers encoded surface form features. A similar mechanism was expected to happen in highway layers in the CBHL encoder in our experiment, encoding pronunciations that were not discernible from orthography. Further research is required to generalize the conclusion by investigating other languages, corpora, and encoder structures.

\begin{figure}[t]
	\centering
		\includegraphics[width=1\columnwidth]{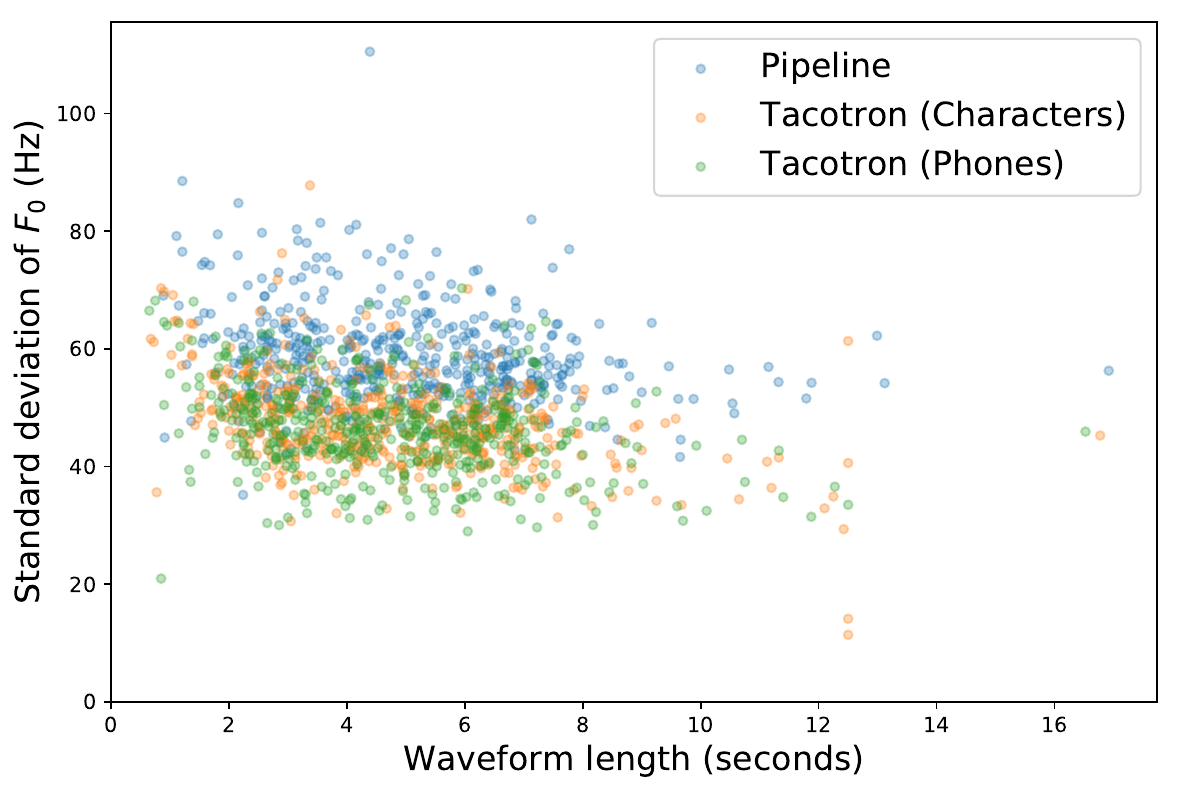}
	\caption{Standard deviation of $F_0$ against output waveform length for samples predicted from our Tacotron systems and pipeline system.}
	\label{fig:stddev-f0}
\end{figure}

The reason English Tacotron systems were outperformed by the pipeline system unlike the Japanese Tacotron systems is unnatural prosody. 
Similar problems are reported from a few studies using Tacotron based systems. According to \citet{Shen2017}, unnatural prosody including unnatural pitch was the most common error in their manual analysis for 100 challenging English sentences. 
We investigated samples from the two best Tacotron systems: self-attention Tacotron using characters and self-attention Tacotron using phones. We listened to their samples that had relatively low average MOS whereas corresponding samples from the pipeline had high MOS of over 4.5. We found that the samples from the Tacotron systems had relatively flat pitch change, which resulted in unnatural prosody. Indeed, the average standard deviations of $F_0$ in the voiced region of generated samples in a test set were 48~Hz for self-attention Tacotron using characters, 46~Hz for self-attention Tacotron using phones, and 59~Hz for the pipeline system. Figure \ref{fig:stddev-f0} shows standard deviations of $F_0$ against the output waveform length for samples from our Tacotron systems and the pipeline system. We can clearly see that samples predicted from our Tacotron system have smaller standard deviations than samples predicted from the pipeline across the whole output length. In addition, we found that the low-score samples contain errors such as mispronunciations or wrongly stressed syllables. This would be one reason the English Tacotron systems have lower scores than the pipeline system unlike the Japanese Tacotron systems. 


\section{Conclusion}
\label{sec:conclusion}

In this paper, we investigated under what conditions sequence-to-sequence based text-to-speech (TTS) could work well given simple input such as text or phonemes in Japanese and English, along with comparing them with comparable deep neural network (DNN) based pipeline TTS systems using complex full-context labels. We empowered models of our sequence-to-sequence based TTS methods instead of enriching linguistic features to see how much enforced models could overcome the linguistic feature limitation. 

We upgraded configurations for our Tacotron based methods from our previous research. 
We increased the parameter size of models for our Tacotron methods. In addition, we tested a convolutional neural network (CNN) based encoder from Tacotron2, along with the CBHL (convolution banks, highway networks and long short-term memory) encoder from the original Tacotron for our TTS systems. Unlike other studies, our baseline pipeline systems used autoregressive probabilistic modeling and a neural vocoder as the sequence-to-sequence based methods do, so the differences in the two methods were mainly about framework approaches.

Our experiment showed that increasing parameter size enabled the sequence-to-sequence based methods using phonemes and accentual-type labels as inputs to outperform the comparable pipeline systems in Japanese. This suggested that a powerful sequence-to-sequence TTS model could learn to compensate for a lack of rich linguistic features. We further investigated the upgraded TTS systems using characters as input in English. We found that the CBHL encoder could learn to disambiguate pronunciation ambiguities given characters as well as phone input better than the CNN encoder. 
However, we also observe that the CBHL encoder could not learn English stressed syllables from characters perfectly and hence resulted in flatter fundamental frequency.

Our future work includes improvements of network architectures and exploring a new way for learning supra-segmental features more appropriately \cite{huang2018music}.

\section{Acknowledgments}

This work was partially supported by a JST CREST Grant (JPMJCR18A6, VoicePersonae project), Japan, MEXT KANKENHI Grants (16H06302, 17H04687, 18H04120, 18H04112, and 18KT0051, 19K24371, 19K24373), Japan, and a Google AI Focused Research Awards Program, Japan. The numerical calculations were carried out on the TSUBAME 3.0 supercomputer at the Tokyo Institute of Technology.

The authors thank Dr. Erica Cooper from NII for her useful feedback on our synthetic speech.

\bibliographystyle{elsarticle-num-names}
\bibliography{sample.bib}

\appendix

\section{Pipeline TTS framework}
\label{sec:pipeline}
\subsection{Front-end in pipeline based TTS}
\label{subsec:japanese-tts-pipeline}

All TTS methods in the pipeline framework have a front-end that uses multiple knowledge-based or statistical models to extract  linguistic features from the input text. These linguistic features are usually represented as full-context labels, which encode the segmental (e.g., phoneme identity) and suprasegmental information (e.g., phrase-level intonation) to utter the text. For example, a typical front-end for English may use a pronunciation dictionary and a decision-tree-based model to conduct the letter-to-sound (or grapheme-to-phoneme) conversion \cite{black1998issues}. It may use other hand-crafted rules or statistic models to infer the phase breaks \cite{taylor1998assigning} and pitch accent \cite{hirschberg1993pitch}. These modules in the front-end are usually language dependent, and the statistical ones require supervised training and hand-annotated datasets.


Note that there are also methods to build a front-end with less language-specific knowledge or supervised training.
For example, the vector space model may be used to derive linguistic features from input text in an unsupervised way \cite{watts2013unsupervised, Lu2013}. However, such a method is not applicable to ideographic languages such as Japanese due to the sparse distribution of written characters.
On the other hand, there are also methods to jointly optimize statistical models in the front-end with the acoustic model in the back-end \cite{Oura2008}. However, this method was only proposed for HMM-based SPSS and cannot be applied to DNN-based SPSS directly. 


\subsection{Back-end in pipeline based TTS}

The back-end in pipeline based TTS has two models: acoustic and waveform. 
The acoustic model has been advanced by replacing the hidden Markov models (HMMs) with DNNs \cite{watts2016hmms}. On the basis of the vanilla feed-forward and recurrent NNs, researchers have further proposed many NN-based acoustic models with improved probabilistic modeling capability.

Particularly, the vanilla feed-forward and recurrent NNs assume statistical independence among the acoustic feature frames \cite{wangxinDARf0}. To amend the independence assumption, autoregressive NN models were proposed to model the acoustic feature distribution of one frame conditioned on the previous frames, which can be implemented by feeding back the previous frames linearly \cite{wangARRMDN} or non-linearly \cite{wangxinDARf0,Lorenzo-Trueba2018}. 
A similar idea can be used to capture the causal dependence between different acoustic feature dimensions \cite{uria2015modelling}.
Meanwhile, unlike the autoregressive NNs, there also other NN-based models using intractable or implicit density functions. Examples of the former case include the post-filtering approach using a restricted Boltzmann machine (RBM) \cite{Chen2015ADeepGenerative} and the trajectory DNN \cite{hashimoto2016trajectory}. The latter case is the generative adversarial network (GAN) \cite{kaneko2017generative}. All the recently proposed NN-based acoustic models alleviate the over-smoothing effect in the vanilla NNs.

There are two types of waveform model in back-end: algorithmic method or neural method. 
Conventional vocoder is an algorithmic method that has long been a core component of SPSS. However, the artifacts in generated waveforms from the vocoder limit the quality of SPSS, which motivates the development of waveform generation methods such as Griffin–Lim \cite{griffin1984signal} in vocoder-free TTS and the neural waveform models \cite{Tamamori2017}. Compared with the Griffin-Lim and related algorithms, the neural waveform models show better performance \cite{XinWang2018}. They are also more flexible as they can work on various types of input acoustic features including mel-spectrogram and vocoder parameters, be implemented with different types of NNs such as GAN \cite{juvela2019gelp} and normalizing flow \cite{prenger2018waveglow}, and incorporate classical signal processing algorithms like linear prediction coefficients (LPC) \cite{valin2018lpcnet} and harmonic-plus-noise model \cite{wang2019}.

\subsection{Pipeline vs sequence-to-sequence TTS}




The DNN-based pipeline and sequence-to-sequence based TTS have a few aspects in common. Both methods can benefit greatly from autoregressive modeling of the acoustic features sequences \cite{XinWang2018, Watts2019} and neural waveform generators
\cite{XinWang2018, DBLP:conf/iclr/PingPGAKNRM18, Shen2017}.
Although it is agreed that sequence-to-sequence based TTS can generally generate more natural speech than traditional methods such as pipeline SPSS, few studies have fully investigated this under the same conditions of density modeling and a waveform generation method. For example, Tacotron \cite{Wang2017} and Tacotron2 \cite{Shen2017} were evaluated against a HMM-based unit selection system \cite{DBLP:conf/interspeech/GonzalvoTCBGS16} and a DNN-based pipeline system without either a neural waveform generator or an autoregressive acoustic model \cite{Heiga16}. 
Some recent studies included a DNN-based pipeline using neural waveform generators, but the non-autoregressive acoustic models still affect the fairness of the comparison \cite{DBLP:conf/nips/GibianskyADMPPR17, DBLP:conf/iclr/PingPGAKNRM18}.


Therefore, if we want to compare pipeline based TTS and sequence-to-sequence based TTS from a framework point of view, it is desirable to select comparable pipeline systems that use an autoregressive decoder with the same waveform generation method as sequence-to-sequence based TTS. Otherwise, differences in the experimental results for the two different approaches can be expected to mainly be caused by the difference in waveform generation techniques and the assumption of probabilistic modeling.

The largest scale experiment to compare various TTS systems including  sequence-to-sequence based methods and pipeline methods is perhaps Blizzard challenge 2019 \cite{Blizzard2019}, in which the task is single-speaker Chinese TTS. Among 24 participating systems, at least 14 were sequence-to-sequence based TTS methods, 5 were SPSS methods, and 2 were unit selection. In addition, one system was natural speech and one was Merlin benchmark \cite{Wu+2016}. The challenge did not intend fair comparison, because the participants could use various methods of waveform generation and even manual labeling or data argumentation with other corpora. Nevertheless, sequence-to-sequence based TTS systems using a neural vocoder obtained relatively high ranks, and the top score was achieved by a DNN-based SPSS system using a neural vocoder, BERT (Bidirectional Encoder Representations from Transformers) based front-end, autoregressive duration model, and non-autoregressive acoustic model with a GAN post-filter\cite{Jiang2019}.

\end{document}